\pdfoutput=1
\documentclass{article}

\title{Trading-profit attribution for the size factor}
\author{Vassilios Papathanakos\footnote{Email: VPapathanakos@intechjanus.com}}
\usepackage{amsmath, graphicx}
\newcommand\Rfig[1]{Figure~\ref{fig:#1}}
\newcommand\Rtab[1]{Table~\ref{tab:#1}}

\begin{document}
\maketitle

\section{Introduction}

In 2015, INTECH Investment Management LLC introduced a novel\footnote{Patent pending.} approach to attribution that focuses on the estimation of the trading profit captured through systematic rebalancing \cite{RP}. This approach was initially applied to the INTECH strategies, which attempt to outperform their benchmarks through the disciplined and risk-controlled use of rebalancing. However, the applicability of this approach extends to a much broader family of portfolios, potentially including most diversified strategies exhibiting regular reconstitution and rebalancing.

In this write-up, we extend the trading-profit attribution methodology to analyze the size factor via simulating equal-weighted portfolios. Equal-weighted portfolios are selected for this purpose because they combine a natural exposure to size with a simpler understanding in terms of the framework of Stochastic Portfolio Theory, so that they furnish a natural test subject for the attribution algorithm.

We investigate the effects of varying the number of securities included in the equal-weighted portfolios, the domicile of these securities (U.S., other developed markets, or emerging markets), transaction costs, and the frequency of the rebalancing. In all cases, we conclude that the trading-profit attribution represents faithfully the long-term outperformance of the equal-weighted portfolio relative to the broad market, demonstrating once more that the size premium is due to volatility capture, rather than a stock-specific factor premium~\cite{BPW}.

This is a technical report focused on describing the details of the experiment -- for more information on the underlying framework and recent work, consult the academic publications \cite{F} and \cite{FK}, the white papers on the INTECH website, or contact the author via email.

\section{Data and methodology}

\subsection{Data sources and investable universes}

\begin{figure}[!htbp]
  \centering
  \includegraphics[width=0.49\textwidth]{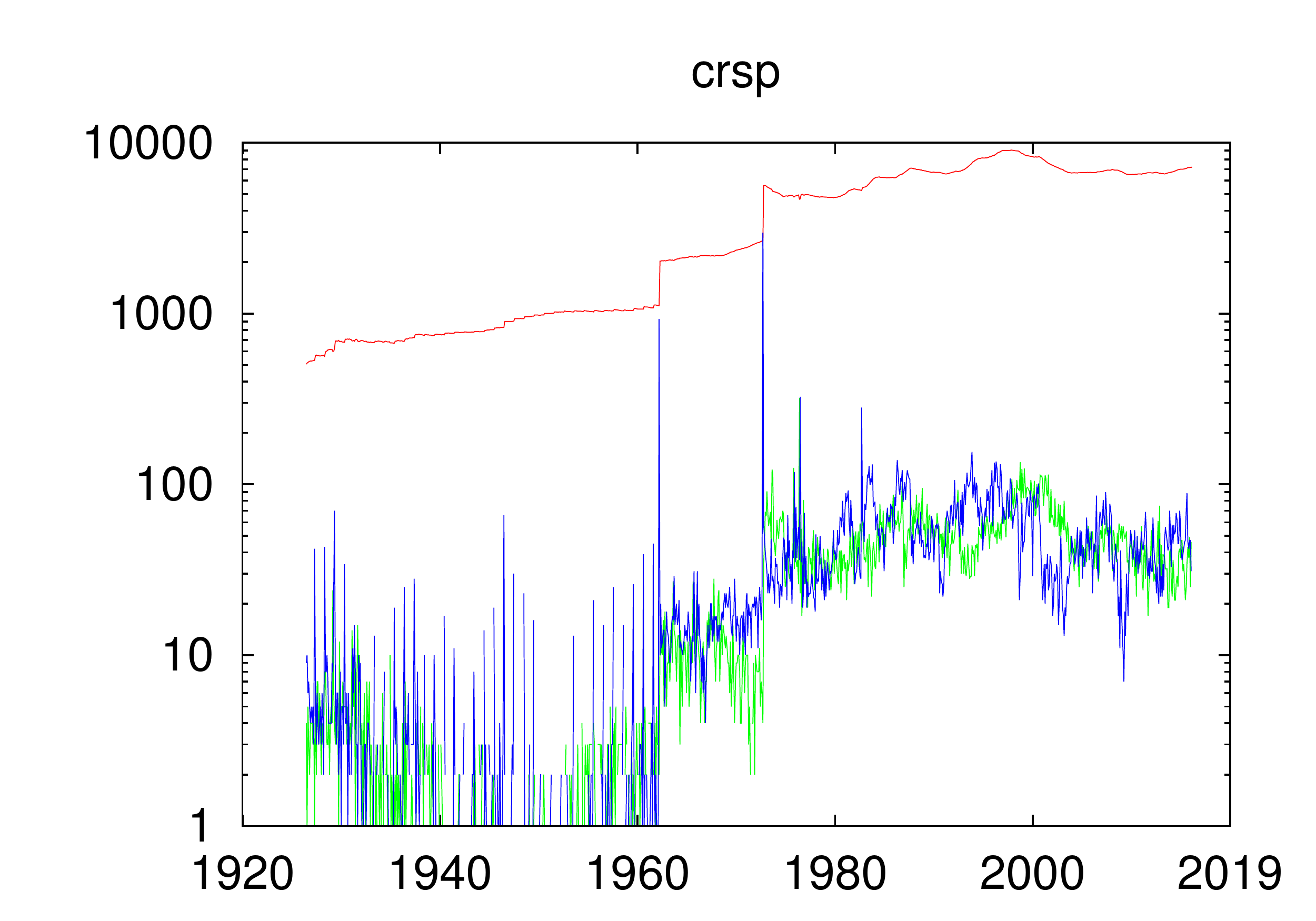}\includegraphics[width=0.49\textwidth]{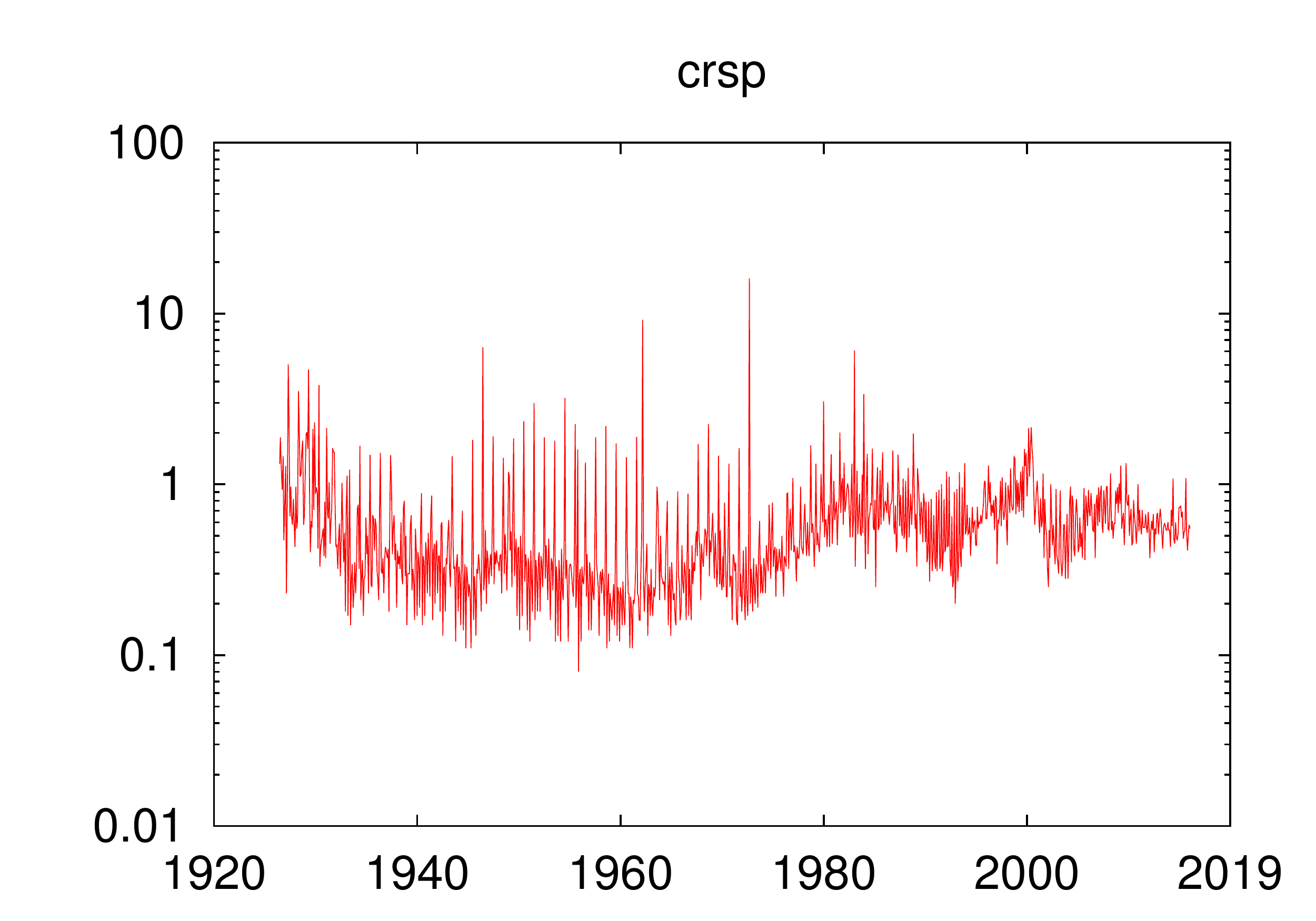}
  \includegraphics[width=0.49\textwidth]{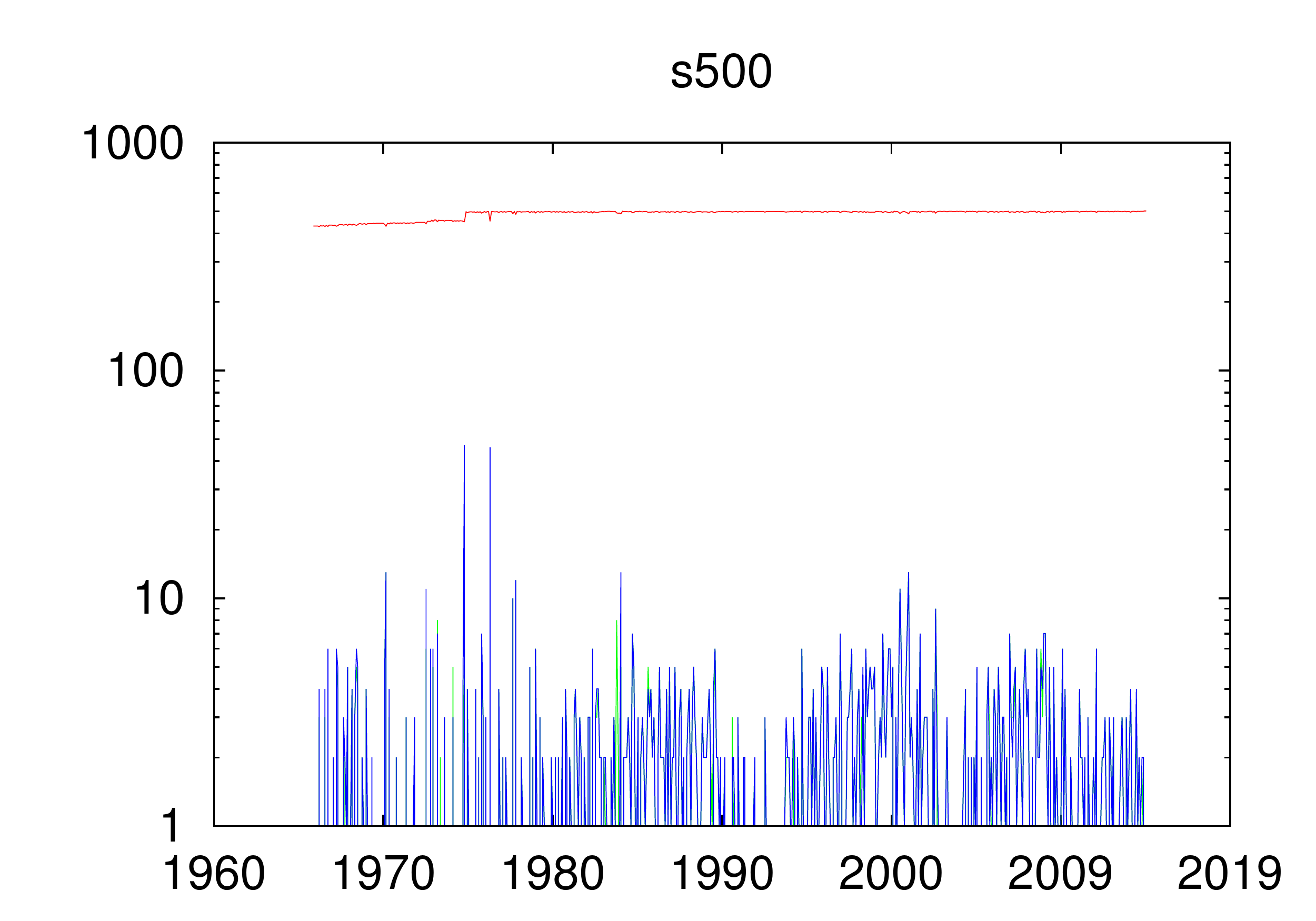}\includegraphics[width=0.49\textwidth]{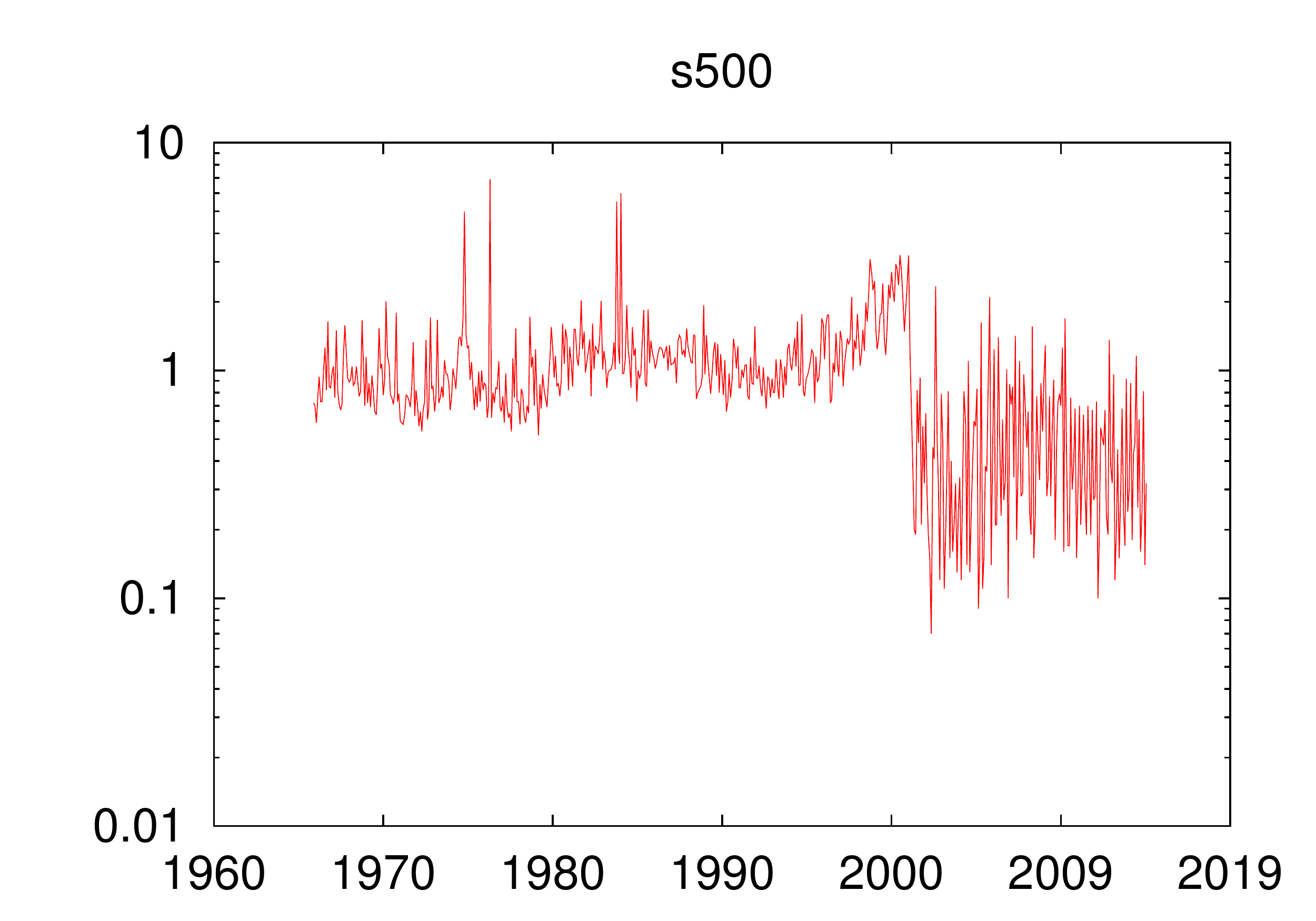}
  \includegraphics[width=0.49\textwidth]{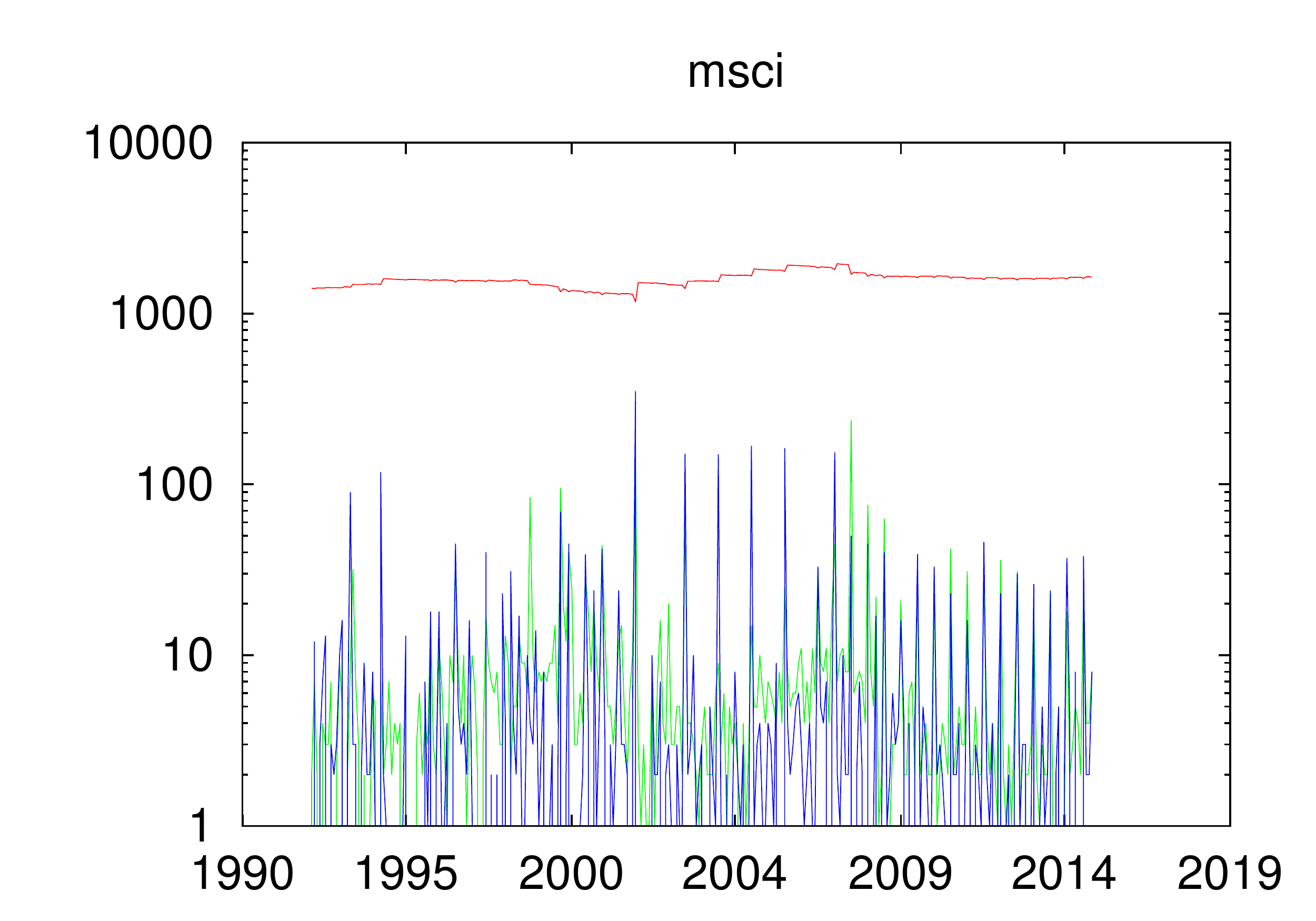}\includegraphics[width=0.49\textwidth]{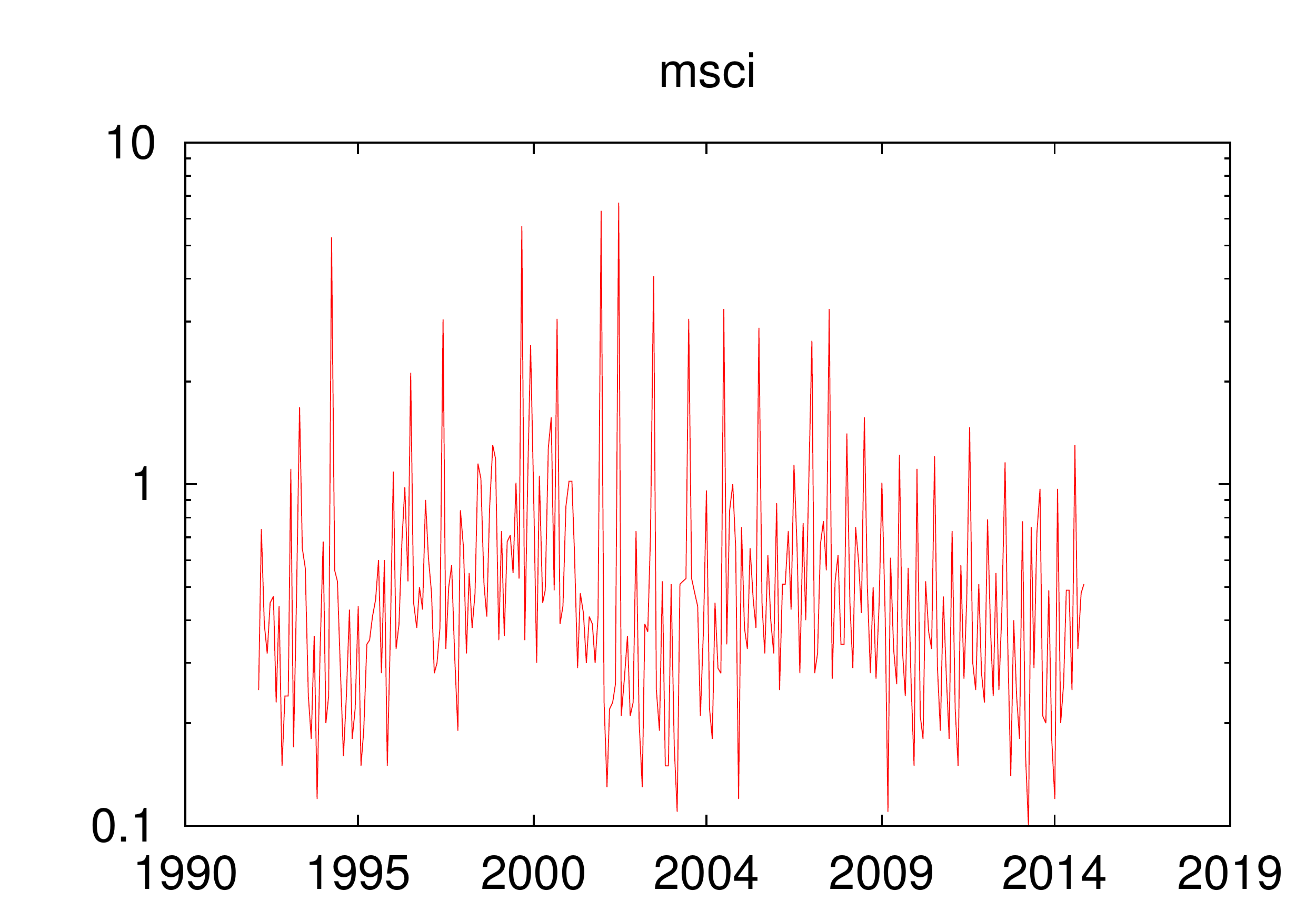}
  \includegraphics[width=0.49\textwidth]{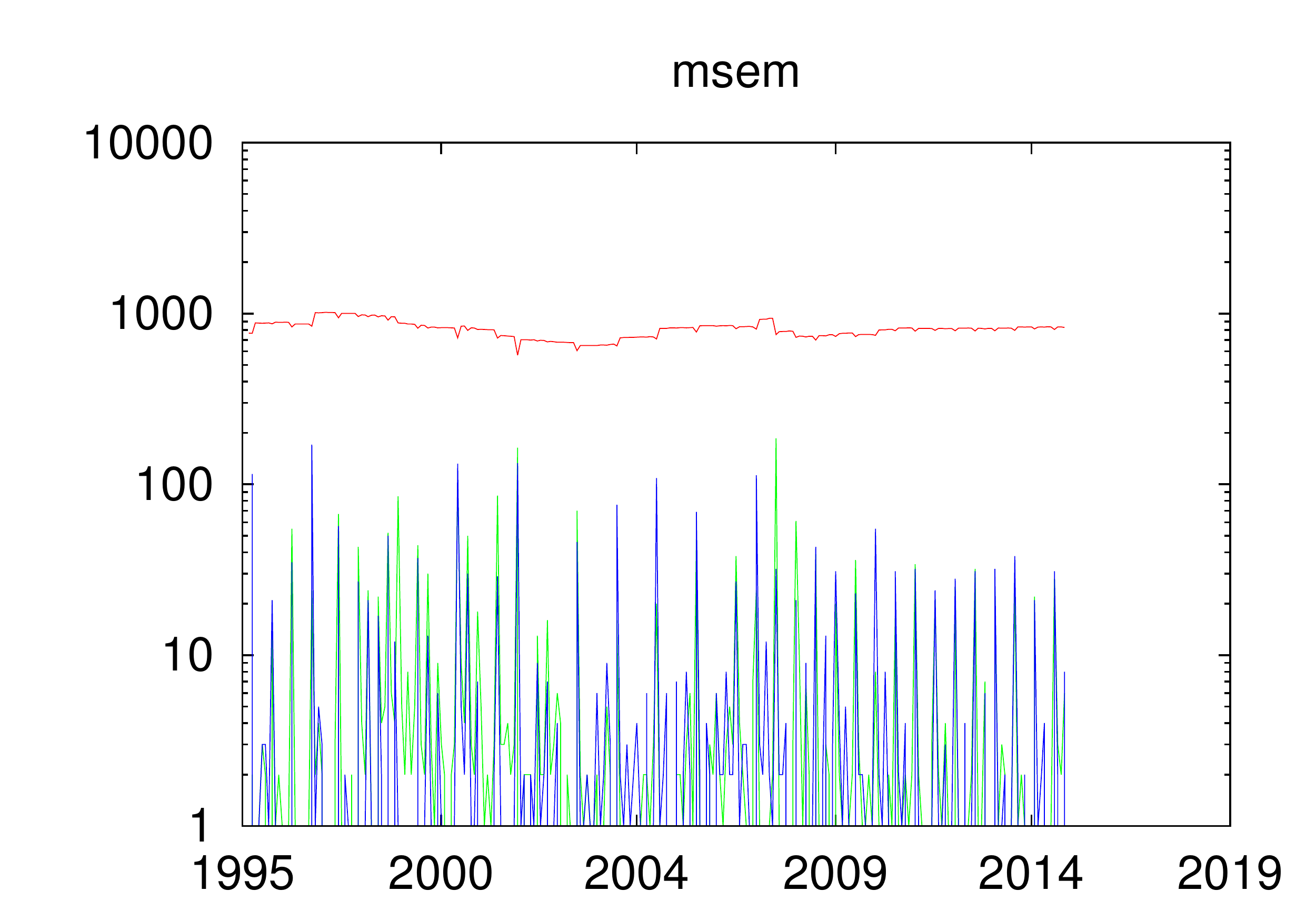}\includegraphics[width=0.49\textwidth]{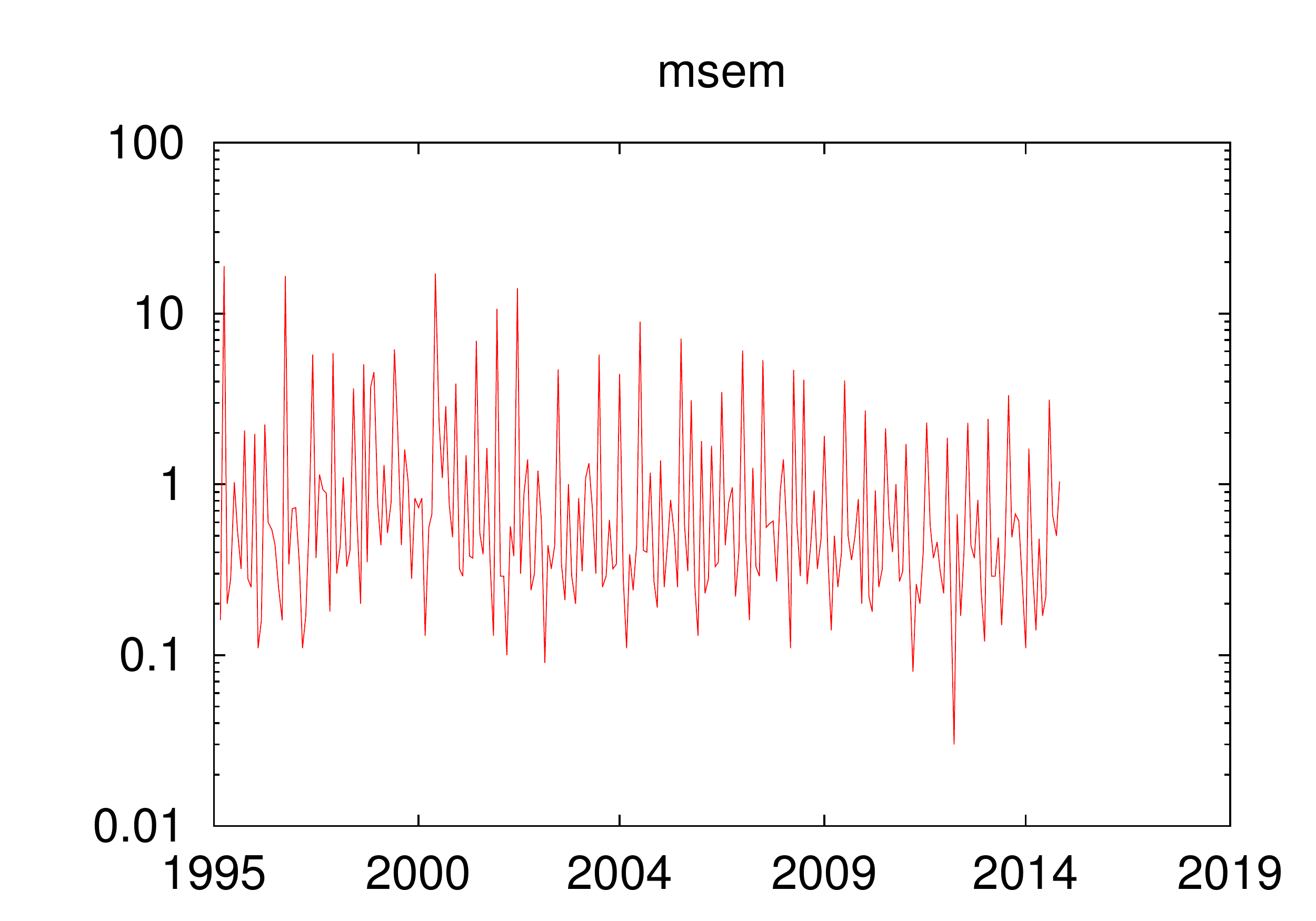}
  \caption{Left: number of stocks that stay (red), leave (green), or enter (blue) the universe at each monthly reconstitution. Right: turnover (\%) for maintaining the broad market index at each monthly reconstitution.}
  \label{fig:recon}
\end{figure}

We simulate portfolios in four collections of securities, or universes, (\texttt{crsp}, \texttt{s500}, \texttt{msci}, \texttt{msem}) described below. In all cases, we reconstitute these universes on the first trading day of each month (cf.~\Rfig{recon} on the following page).

\subsubsection{\texttt{crsp}}

This universe consists of securities contained in the daily stock database of the Center for Research in Securities Prices \cite{C}. We do not include stocks that are not traded on the New York Stock Exchange (NYSE), the American Stock Exchange (AMEX), the NASDAQ stock market, and the Arca exchange. The data span the period January 1927 through December 2015.

\subsubsection{\texttt{s500}}

This universe consists of securities contained in the S\&P~500 Index \cite{S}. The data span the period January 1966 through December 2015.

\subsubsection{\texttt{msci}}

This universe consists of securities contained in the MSCI World Index, which includes securities from the U.S.~and other developed markets \cite{M}. The data span the period January 1992 through December 2015, and 25 countries appear overall.

\subsubsection{\texttt{msem}}

This universe consists of securities contained in the MSCI Emerging Markets Index, which includes securities from the U.S.~and other developed markets \cite{M}. The data span the period January 1995 through December 2015, and 31 countries appear overall.

\subsection{Construction of the equal-weighted portfolios}

\begin{figure}[!htbp]
  \centering
  \includegraphics[width=0.49\textwidth]{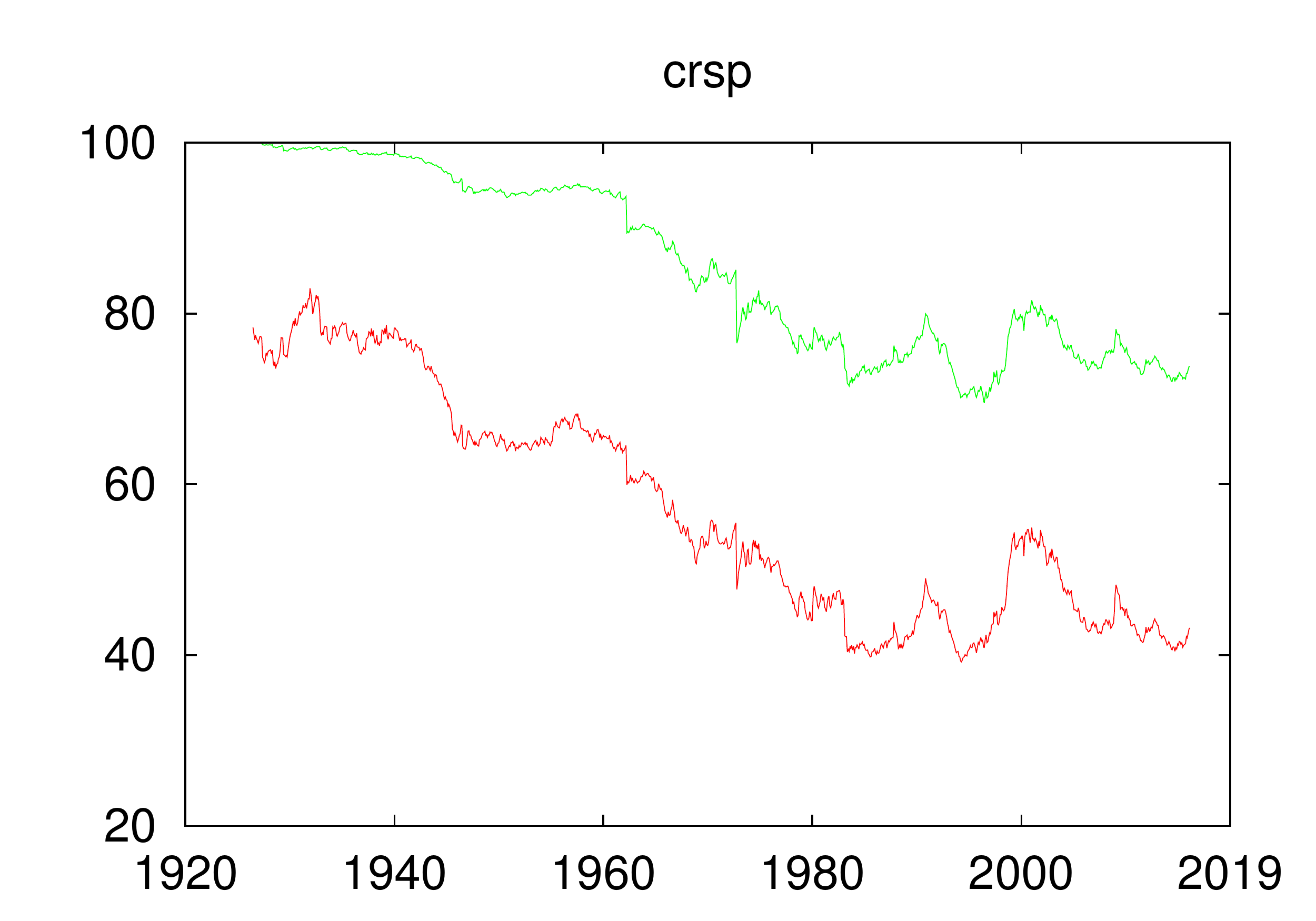}
  \includegraphics[width=0.49\textwidth]{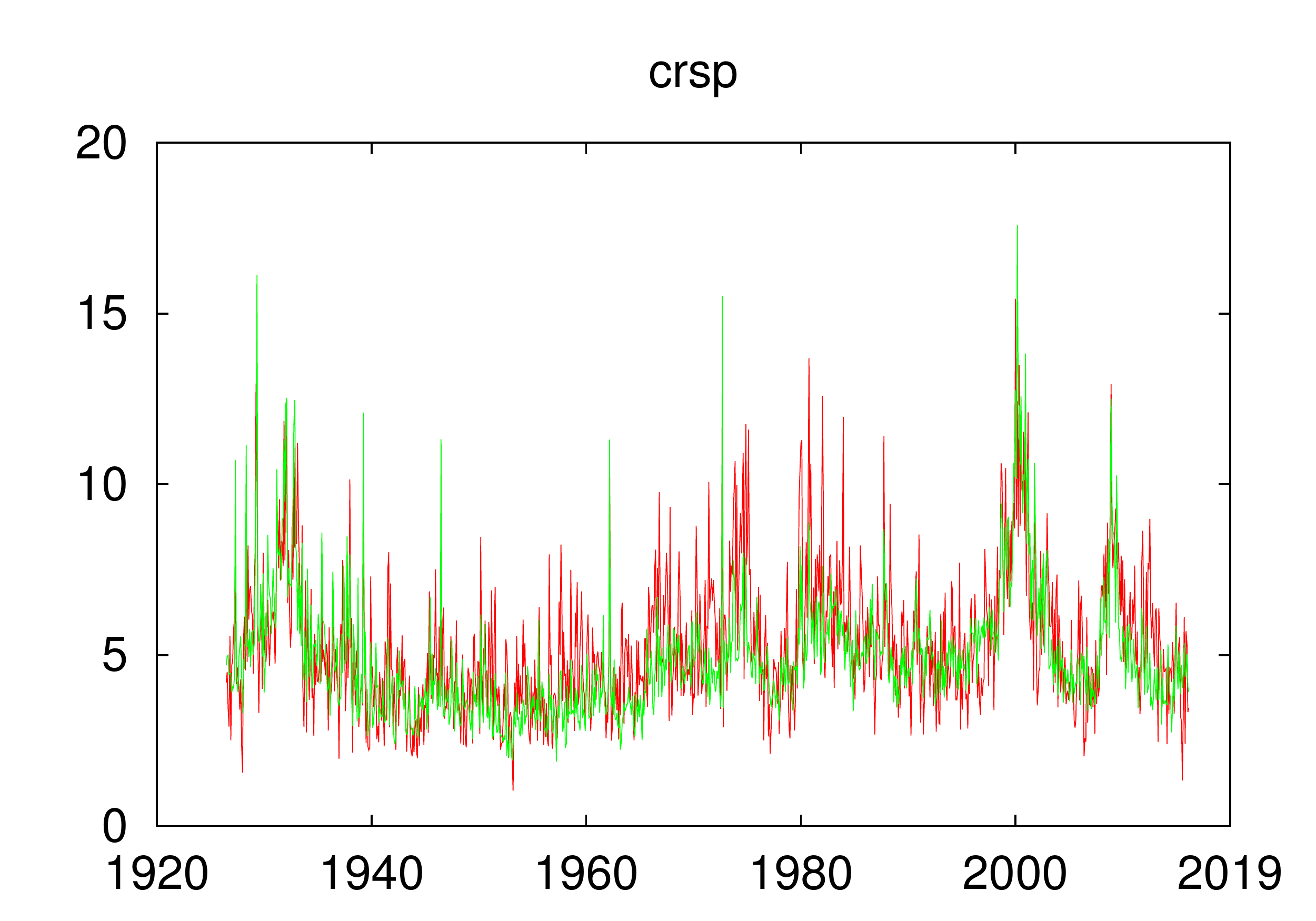}\\
  \includegraphics[width=0.49\textwidth]{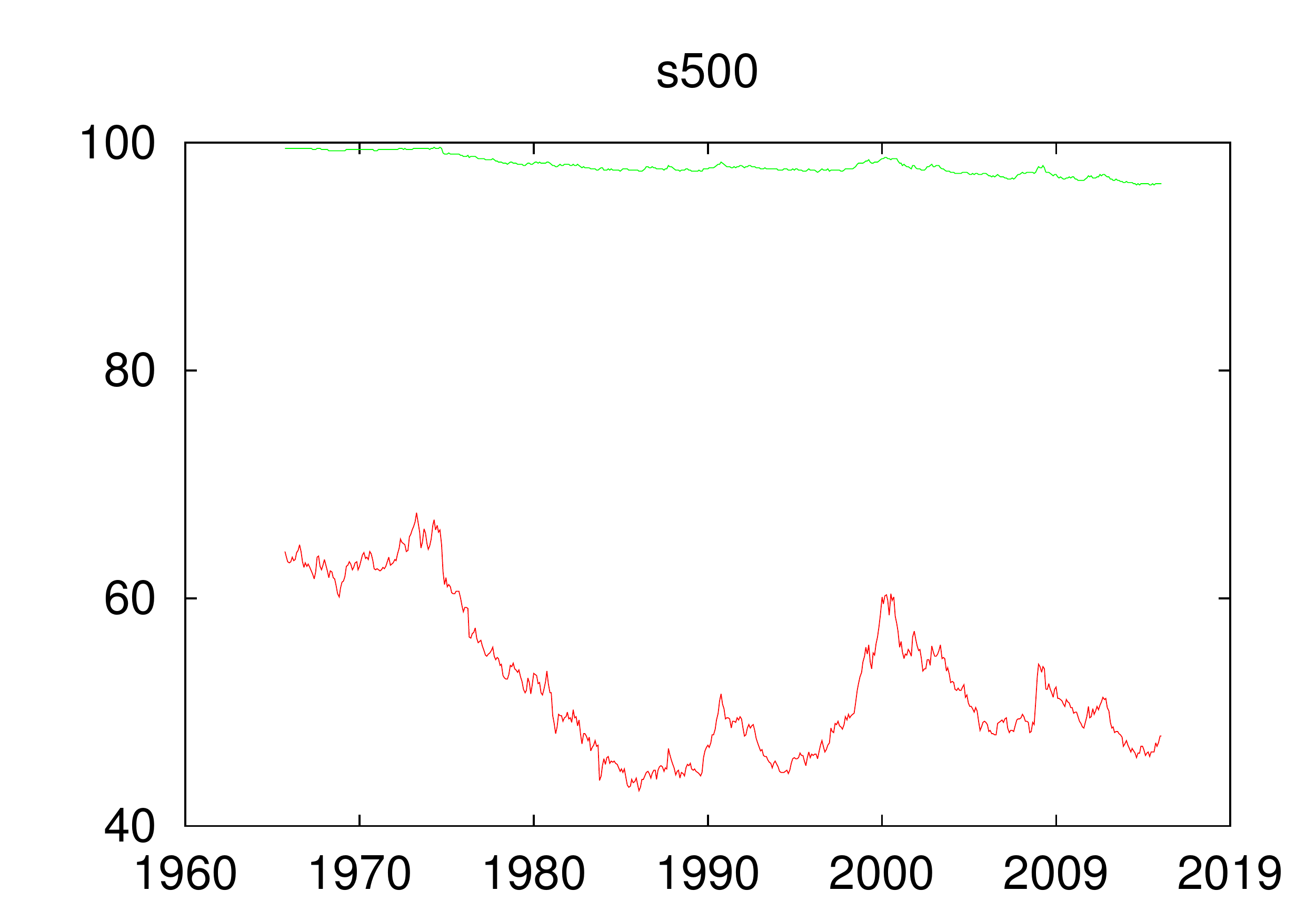}
  \includegraphics[width=0.49\textwidth]{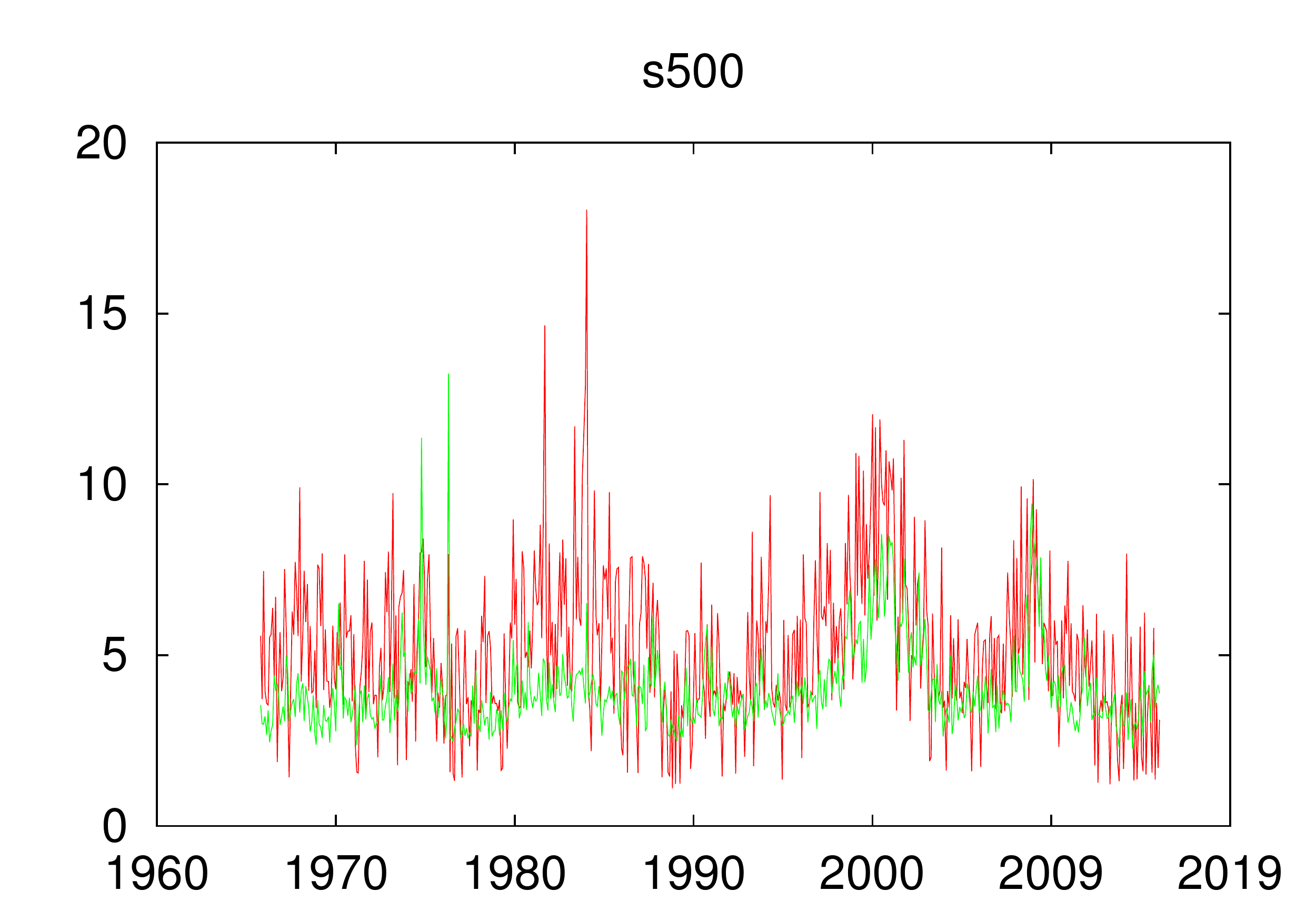}\\
  \includegraphics[width=0.49\textwidth]{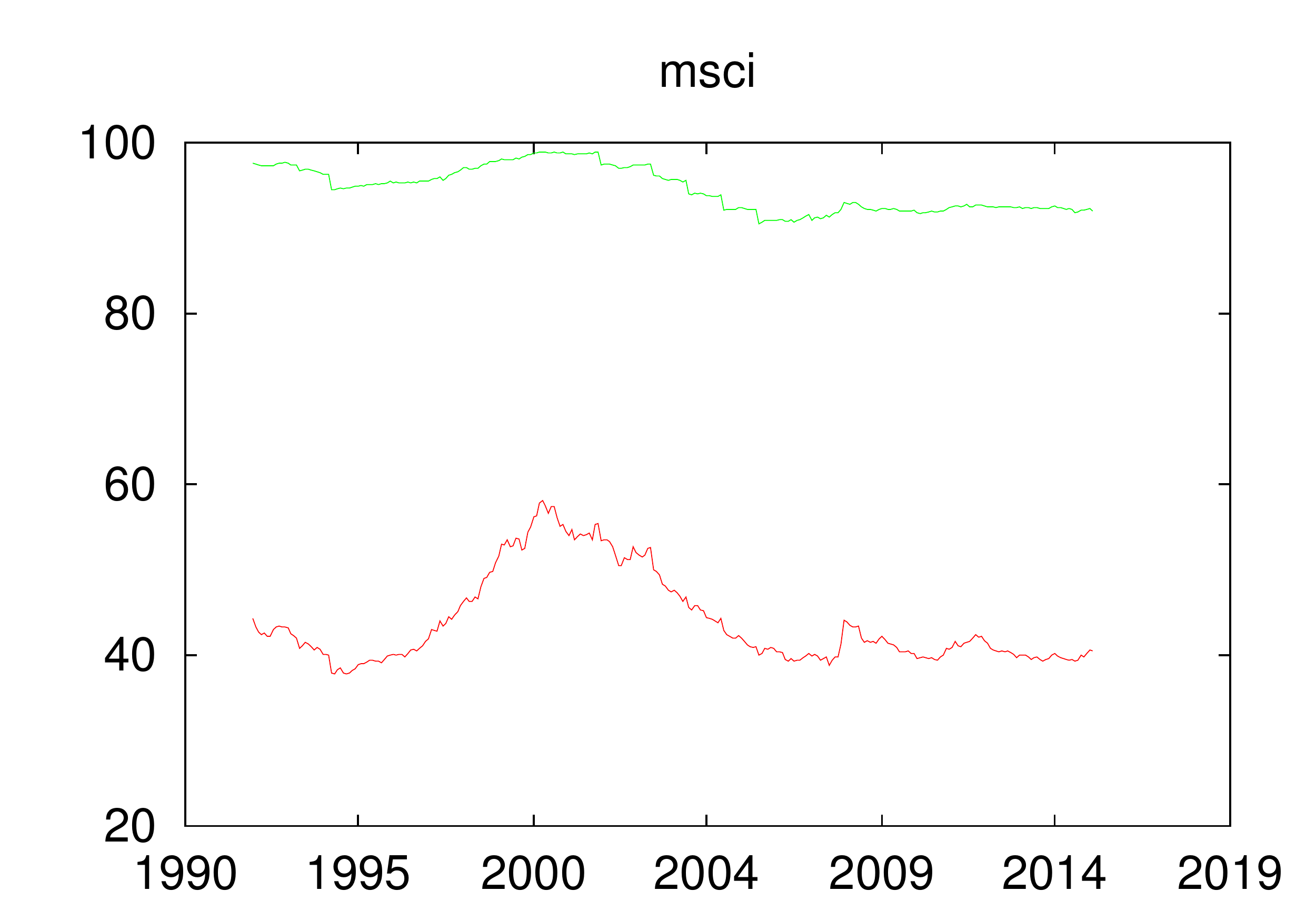}
  \includegraphics[width=0.49\textwidth]{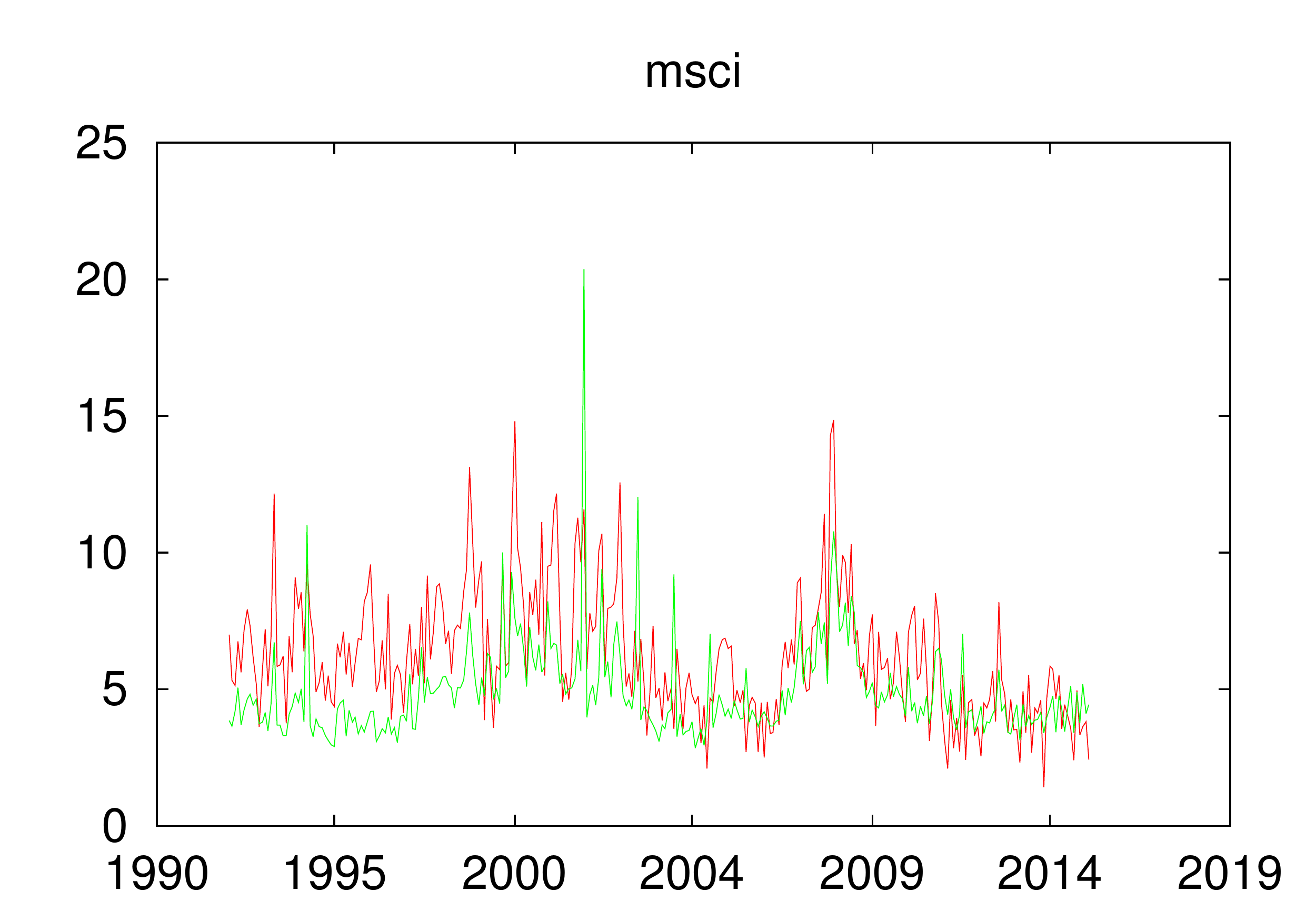}
  \includegraphics[width=0.49\textwidth]{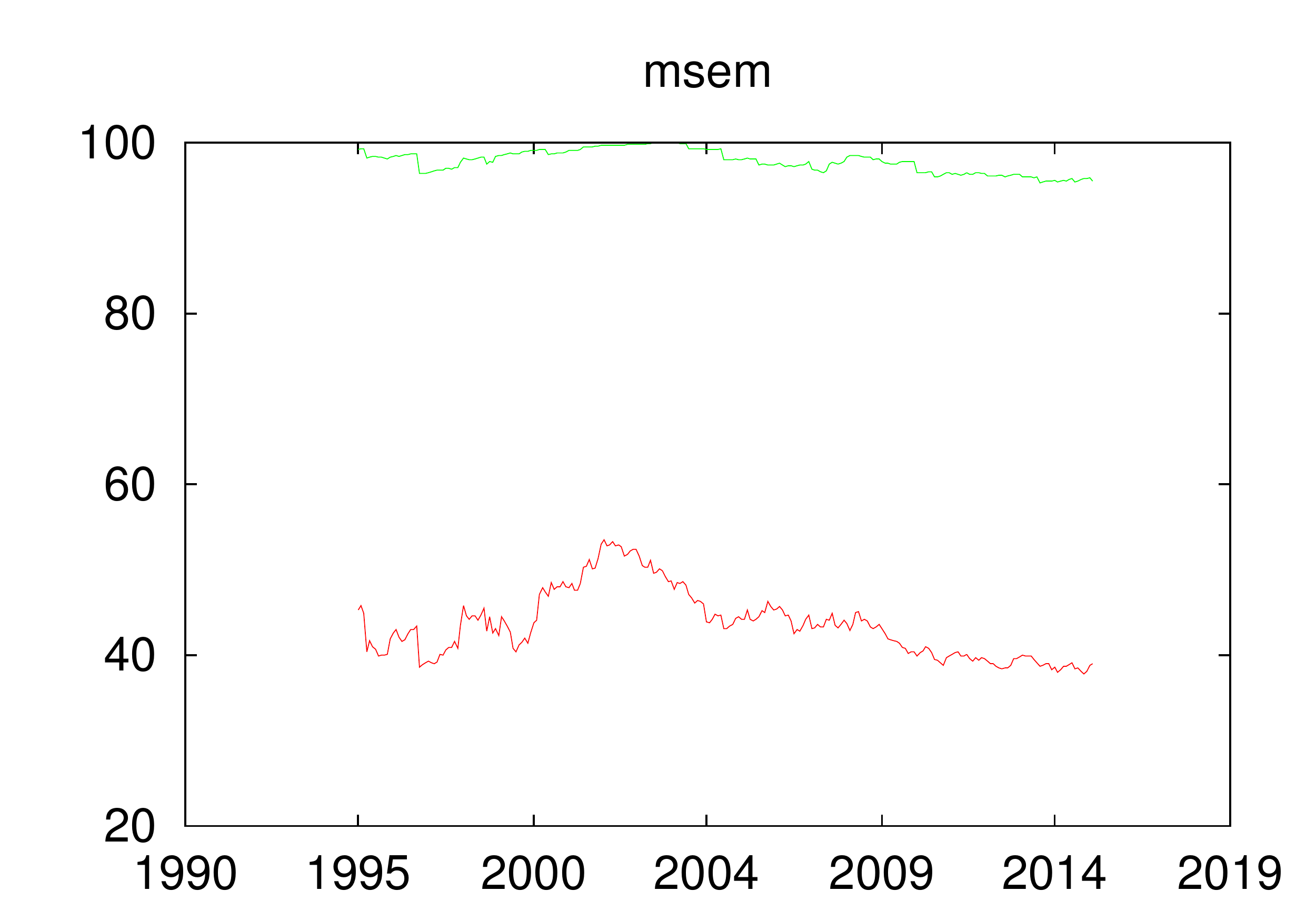}
  \includegraphics[width=0.49\textwidth]{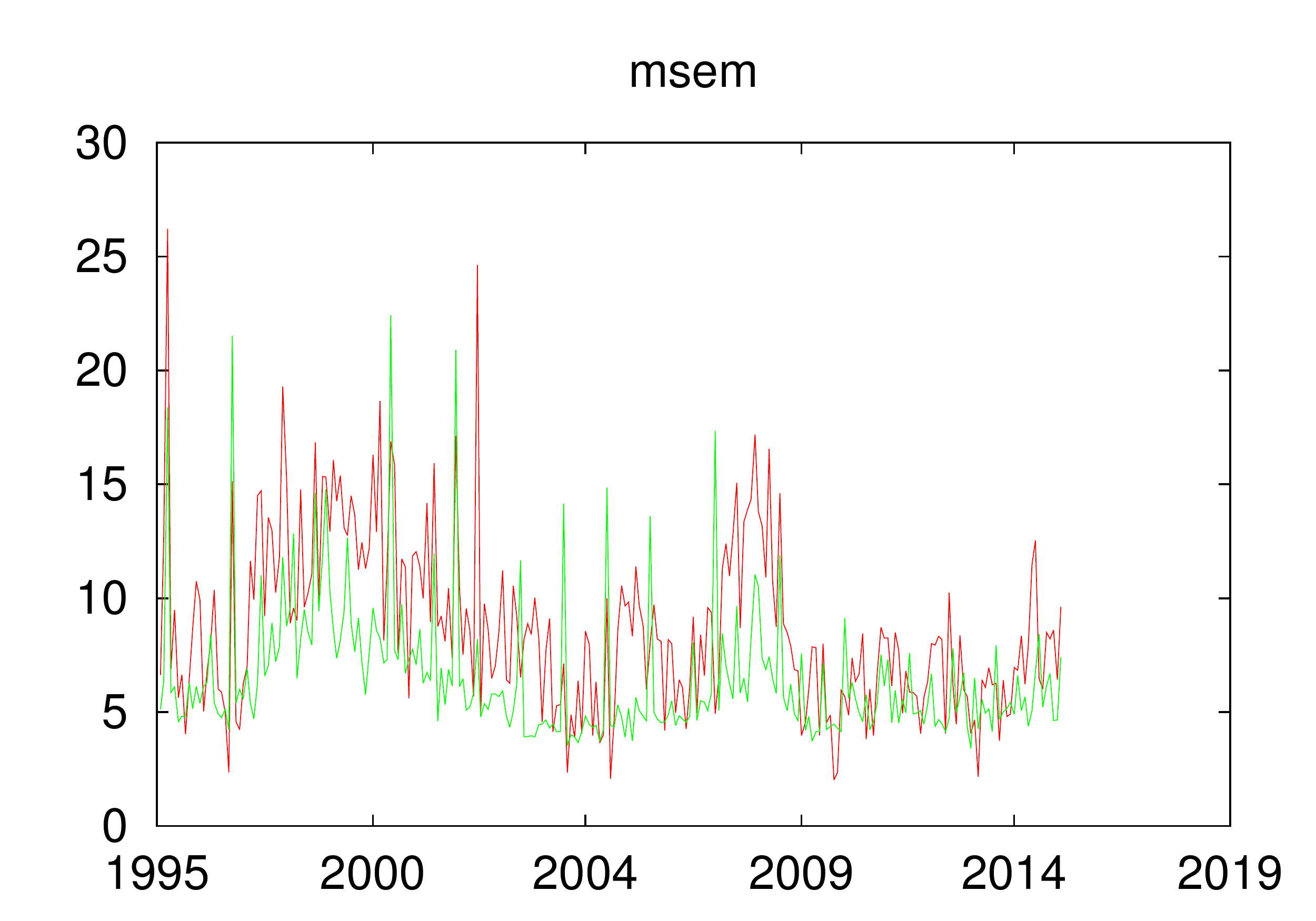}
  \caption{Left: Percentage of the market capitalization in the top $n$ stocks (red: \texttt{lrg}; green: \texttt{sml}). Right: Monthly turnover of the equal-weighted portfolio consisting of the top $n$ stocks (red: \texttt{lrg}; green: \texttt{sml}).}
  \label{fig:frc}
\end{figure}

At the start of each month, we simulate a rebalancing trade that results in an equal-weighting portfolio on the top $n$ names in the universe, ranked by market capitalization; $n$ is chosen below to take two different values, denoted \texttt{lrg} and \texttt{sml} (cf.~\Rtab{Val}). These values are chosen to ensure that sufficient securities appear in the universe throughout the simulation period, as well as to highlight the effect of rebalancing in the larger-capitalization versus the smaller-capitalization securities (cf.~\Rfig{frc}).

\begin{table}[!hbtp]
  \centering
  \begin{tabular}{|c|c|c|}
    \hline
    \textbf{Universe} & \texttt{lrg} & \texttt{sml} \\
    \hline
    \texttt{crsp}     & 100          & 500          \\
    \texttt{s500}     & 50           & 400          \\
    \texttt{msci}     & 100          & 1000         \\
    \texttt{msem}     & 50           & 650          \\
    \hline
  \end{tabular}
  \caption{Values of $n$ used to define the thresholds for inclusion to the equal-weighted portfolios; two options are considered in each universe.}
  \label{tab:Val}
\end{table}

The performance of the equal-weighted portfolio is measured relative to the broad market, defined as the cap-weighted portfolio that consists of all securities in the universe at the beginning of that month. We assume that there are no discretization effects, i.e.,~that all stocks can be held or traded at arbitrary (including fractional) amount of shares.

\subsection{Stochastic Portfolio Theory}

According to the framework of Stochastic Portfolio Theory \cite{F}, the relative performance of an equal-weighted portfolio can be approximately\footnote{This is only an approximation due to complications that arise due to various corporate events (such as mergers and acquisitions), the ambiguity of free float versus total shares outstanding, as well as the existence of dividends.} understood as the sum of three contributions:
\begin{enumerate}
\item the exposure to size, which can be quantified as the change in the average logarithm of the market weight of each security in the equal-weighted portfolio;
\item the rebalancing premium, i.e.,~the trading profit captured through rebalancing to the target weights for those securities that remain in the top $n$ by market cap through consecutive reconstitutions;
\item the leakage effect, i.e.,~the reconstitution drag due to the detrimental selling out of securities because they are no longer in the top $n$ (or, even, the investable universe).
\end{enumerate}
The exposure to size can be estimated very accurately for the equal-weighted portfolio; the leakage effect can also be estimated, albeit with greater uncertainty. This means that equal-weighted portfolios furnish a convenient testing ground for evaluating the accuracy of the novel trading-profit attribution.

A simple and robust estimate of the leakage is given by the expression (cf.~\Rtab{cal})
\begin{multline}
  \text{(leakage)} = \text{(calibration factor)} \times \\
  \left\{\left[\text{(return of the equal-weighted portfolio on top $n$ stocks)}-\right.\vphantom{\int}\right.\\
    \left.\text{(return of the cap-weighted portfolio on top $n$ stocks)}\right]-\\
  \left.\vphantom{\int}\text{(exposure to size)}\right\}
\end{multline}

\begin{table}[!hbtp]
  \centering
  \begin{tabular}{|c|cc|}
    \hline
    \textbf{Universe} & \multicolumn{2}{|c|}{\textbf{Calibration factor}} \\
    & \textbf{\texttt{lrg}} & \textbf{\texttt{sml}}     \\
    \hline
    \texttt{crsp}     & 0.3                   & 0.3                       \\
    \texttt{s500}     & 0.45                  & 0.55                      \\
    \texttt{msci}     & 0.45                  & 0.55                      \\
    \texttt{msem}     & 0.6                   & 0.65                      \\
    \hline
  \end{tabular}
  \caption{Calibration factors for leakage computation used for the estimation of the rebalancing premium.}
  \label{tab:cal}
\end{table}

\subsection{Trading-profit attribution}

We apply the trading-profit attribution methodology as described in \cite{RP} with a single modification: when looping over previous buys, we break out of the loop if we encounter reconstitution buys, i.e.,~buys of stocks from zero initial weight.

This is intuitively motivated by the observation that these trades are triggered not by rebalancing (namely, trading in direct reaction to price movements, in the opposite direction), but by the appearance of the stock to the list of investable securities (the top $n$ stocks in the universe).

This is not a material consideration for most realistic strategies which take steps to avoid excessive trading triggered by reconstitution events. However, in our case, where we examine simple, unmanaged strategies, the effect can be significant and must be explicitly addressed.

\subsection{Transaction costs}

When transaction costs are taken into account, we assume that they equal 40~bps. Also, we reduce the portfolio performance at the time of trade by the product of the transaction cost times the absolute value of the weight change.

In that case, we also adjust the trading profit at the time of a sell by subtracting from it twice the product of the transaction cost times the absolute value of the matched weight change. Finally, the trading profit for that day is further reduced by twice the product of the transaction cost times the absolute value of the unmatched sell weight.


\section{Effect of universe magnitude}

\begin{figure}[!htbp]
  \centering
  \includegraphics[width=0.49\textwidth]{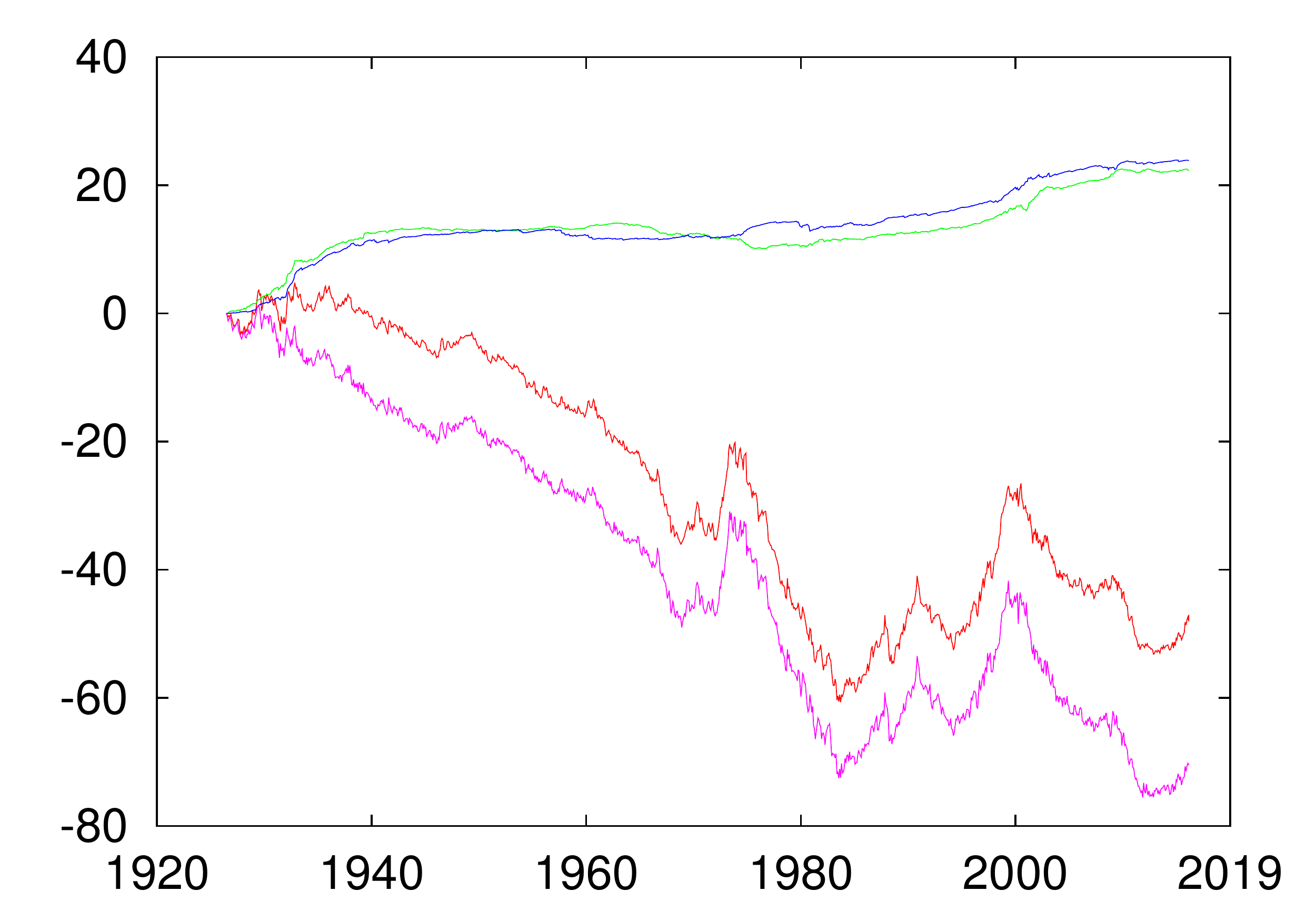}
  \includegraphics[width=0.49\textwidth]{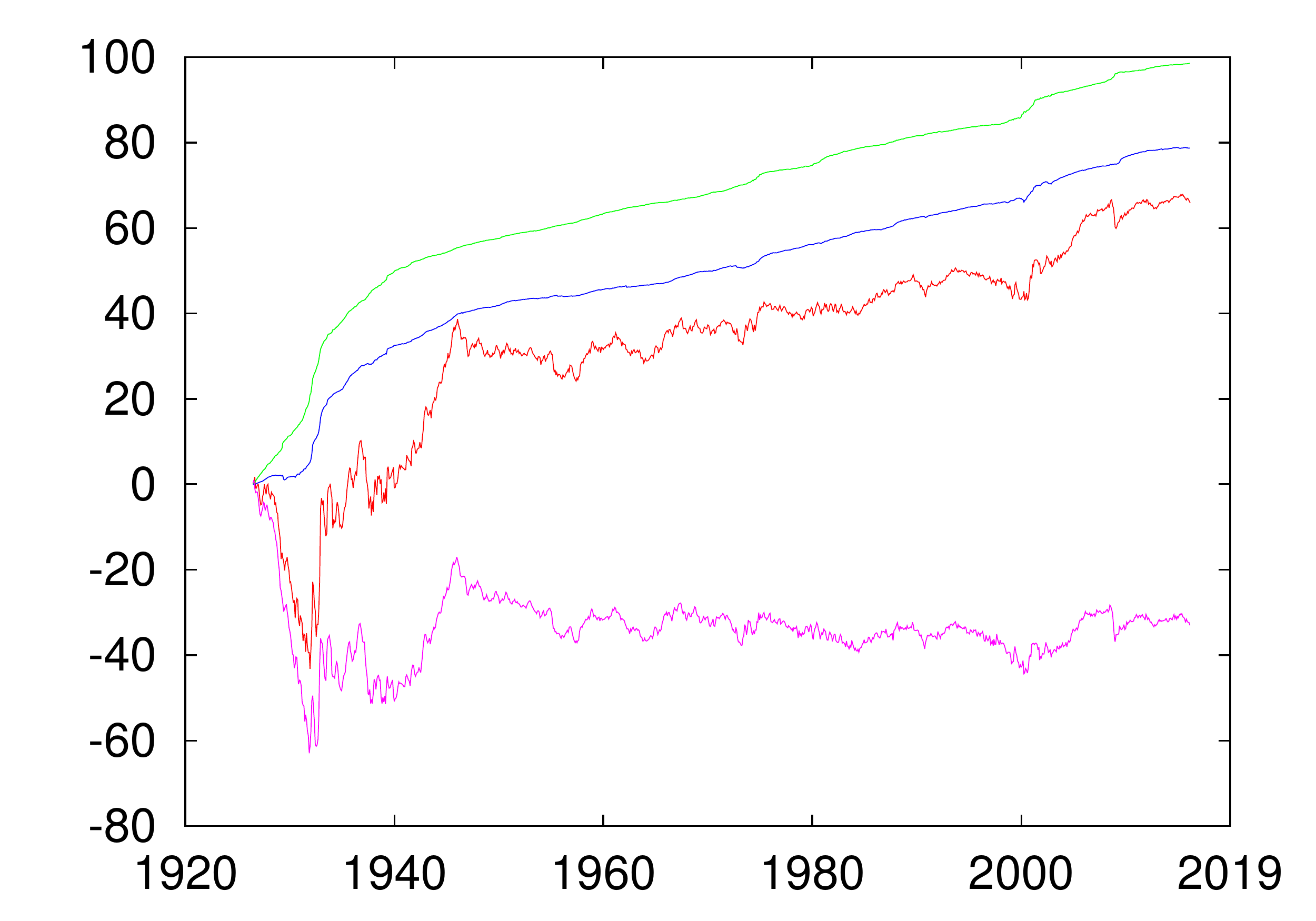}
  \caption{Cumulative performance of the equal-weighted portfolio for the top $n$ stocks (left: \texttt{lrg}; right: \texttt{sml}) relative to the full market (red), the SPT-based estimate of the rebalancing premium (green), the trading-profit attribution (blue), and the size exposure (pink) for the \texttt{crsp} universe.}
  \label{fig:pft-crsp}
\end{figure}

\begin{figure}[!htbp]
  \centering
  \includegraphics[width=0.49\textwidth]{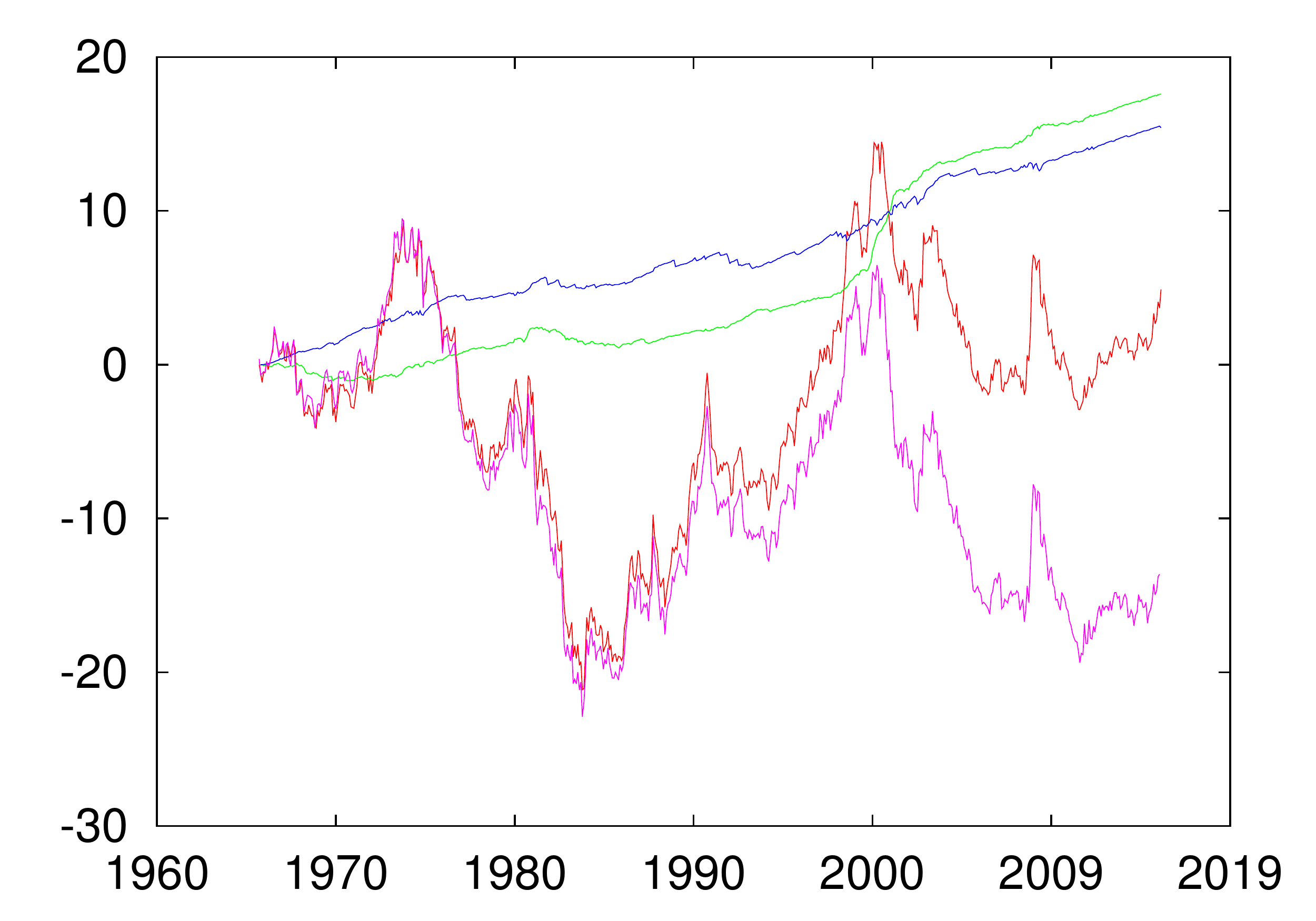}
  \includegraphics[width=0.49\textwidth]{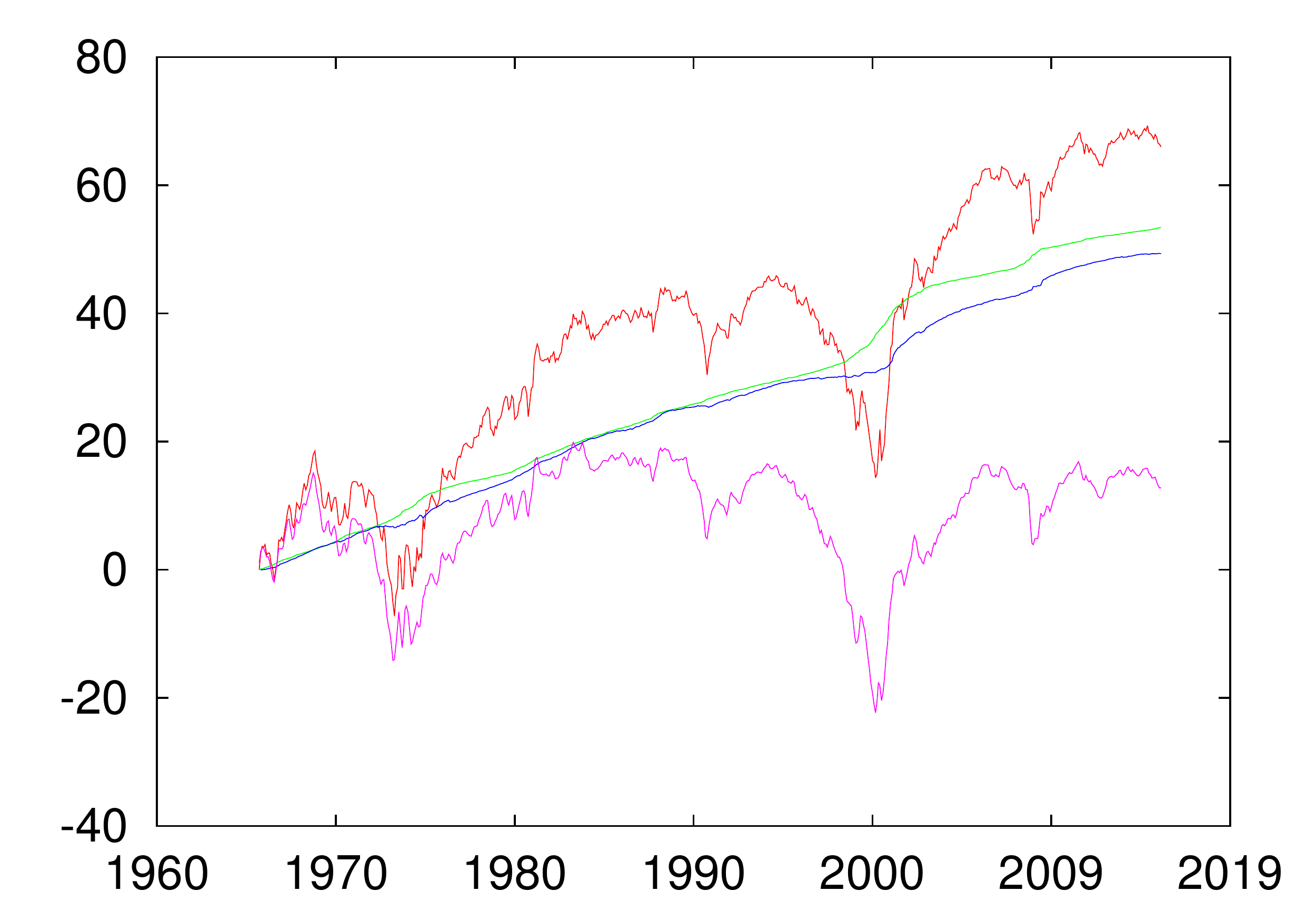}
  \caption{Same as \Rfig{pft-crsp} for the \texttt{s500} universe.}
  \label{fig:pft-s500}
\end{figure}

\begin{figure}[!htbp]
  \centering
  \includegraphics[width=0.49\textwidth]{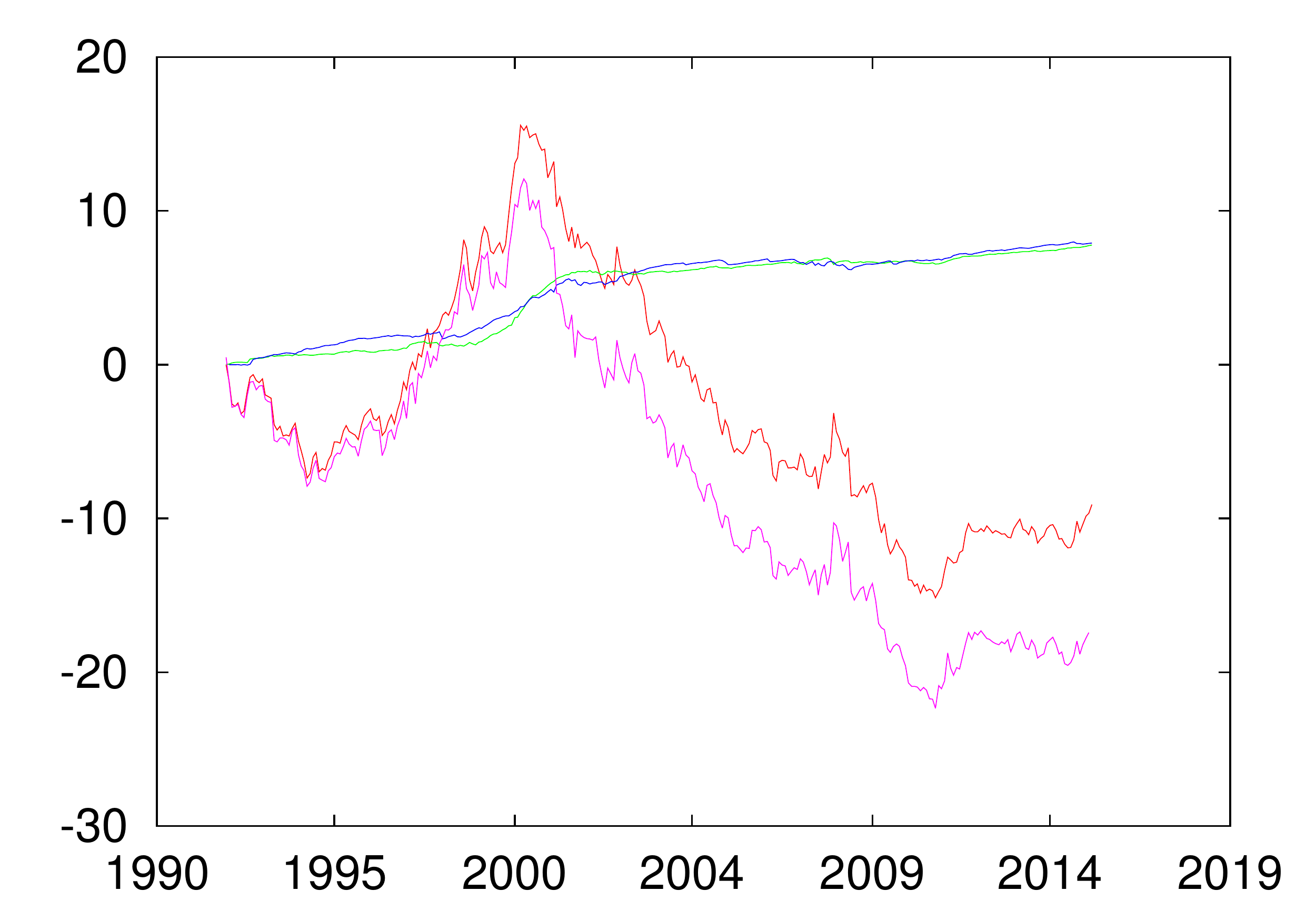}
  \includegraphics[width=0.49\textwidth]{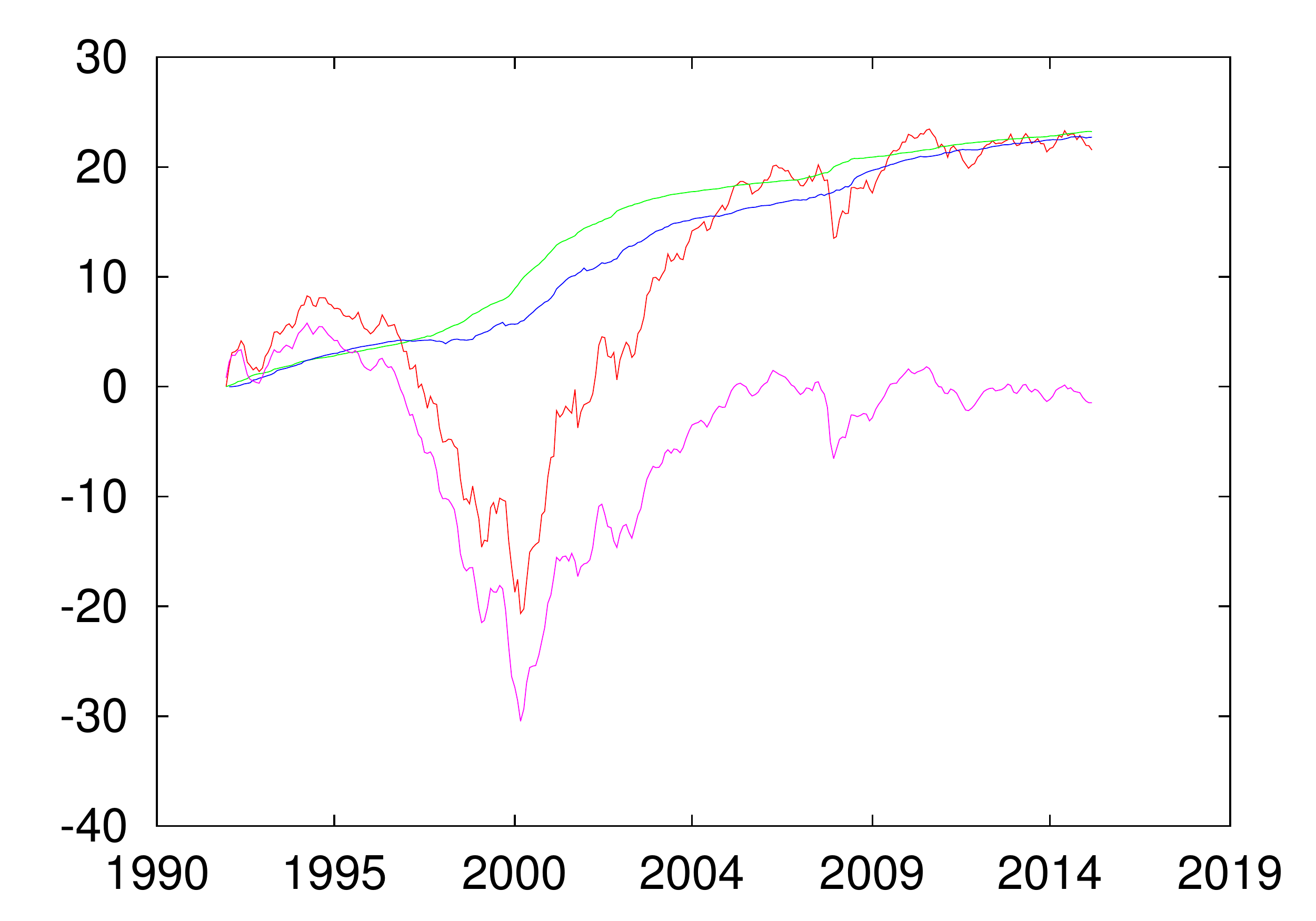}
  \caption{Same as \Rfig{pft-crsp} for the \texttt{msci} universe.}
  \label{fig:pft-msci}
\end{figure}

\begin{figure}[!htbp]
  \centering
  \includegraphics[width=0.49\textwidth]{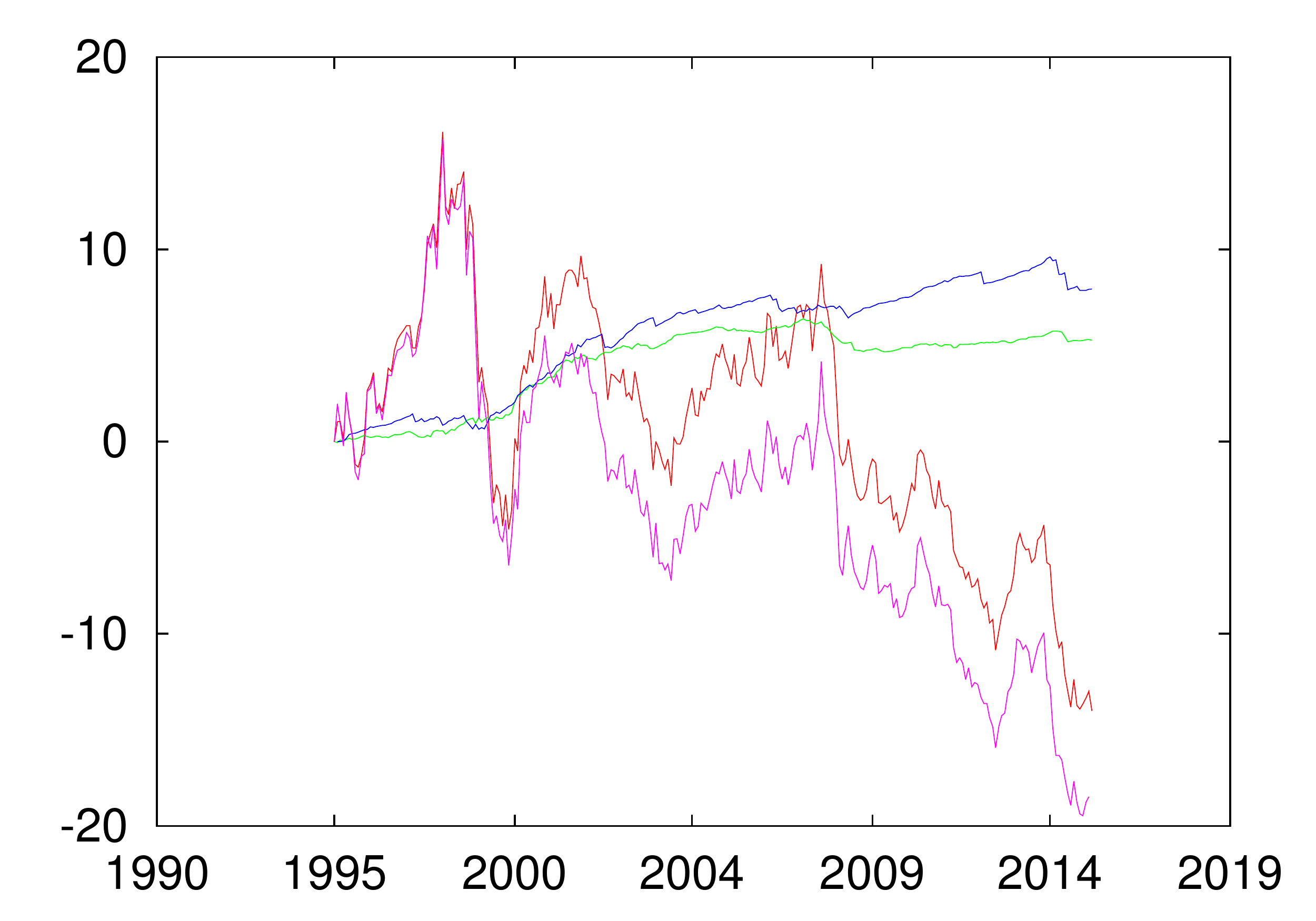}
  \includegraphics[width=0.49\textwidth]{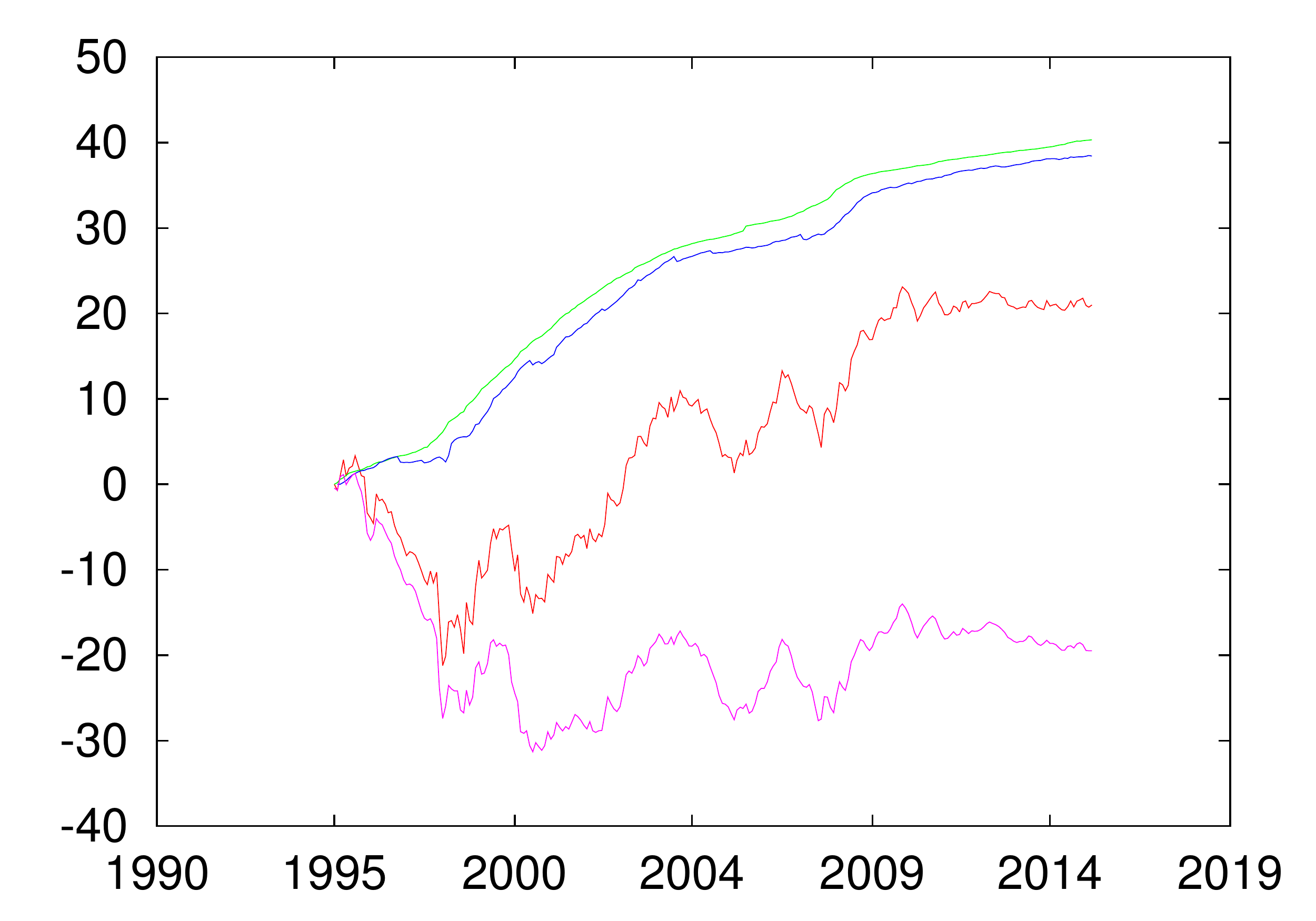}
  \caption{Same as \Rfig{pft-crsp} for the \texttt{msem} universe.}
  \label{fig:pft-msem}
\end{figure}

In this section, we explore how the choice of the size of the equal-weighted portfolio, i.e.,~how many of the top stocks to include, affects the results. For convenience, as explained above, we select two values for this threshold number $n$, denoted by \texttt{lrg} and \texttt{sml} respectively (cf.~\Rtab{Val}). Furthermore, we assume that there are no transaction costs, and that rebalancing occurs on a monthly frequency. The results of the simulations are shown in Tables~\ref{tab:sum-crsp}--\ref{tab:sum-msem}, as well as in Figures~\ref{fig:pft-crsp}--\ref{fig:pft-msem}.

\subsection{\texttt{crsp} universe}

\begin{table}[!hbtp]
  \centering
  \begin{tabular}{|c|cc|cc|}
    \hline
    \textbf{Series} & \multicolumn{2}{|c|}{\texttt{lrg}} & \multicolumn{2}{|c|}{\texttt{sml}} \\
    & \textbf{Mean} & \textbf{St. dev.}  & \textbf{Mean} & \textbf{St. dev.} \\
    \hline
    Equal-weighted relative return &-0.53 & 2.71 & 0.74 & 4.66 \\
    Rebalancing premium            & 0.25 & 0.34 & 1.11 & 0.54 \\
    Trading profit                 & 0.27 & 0.33 & 0.88 & 0.53 \\
    \hline
  \end{tabular}
  \caption{Statistics for the equal-weighted portfolio on the \texttt{crsp} universe.}
  \label{tab:sum-crsp}
\end{table}

\subsection{\texttt{s500} universe}

\begin{table}[!hbtp]
  \centering
  \begin{tabular}{|c|cc|cc|}
    \hline
    \textbf{Series} & \multicolumn{2}{|c|}{\texttt{lrg}} & \multicolumn{2}{|c|}{\texttt{sml}} \\
    & \textbf{Mean} & \textbf{St. dev.}  & \textbf{Mean} & \textbf{St. dev.}  \\
    \hline
    Equal-weighted relative return & 0.10 & 3.21 & 1.32 & 4.56 \\
    Rebalancing premium            & 0.35 & 0.29 & 1.07 & 0.25 \\
    Trading profit                 & 0.31 & 0.32 & 0.99 & 0.31 \\
    \hline
  \end{tabular}
  \caption{Statistics for the equal-weighted portfolio on the \texttt{s500} universe.}
  \label{tab:sum-s500}
\end{table}

\newpage

\subsection{\texttt{msci} universe}

\begin{table}[!hbtp]
  \centering
  \begin{tabular}{|c|cc|cc|}
    \hline
    \textbf{Series} & \multicolumn{2}{|c|}{\texttt{lrg}} & \multicolumn{2}{|c|}{\texttt{sml}} \\
    & \textbf{Mean} & \textbf{St. dev.}  & \textbf{Mean} & \textbf{St. dev.} \\
    \hline
    Equal-weighted relative return &-0.38 & 2.73 & 0.90 & 3.38 \\
    Rebalancing premium            & 0.32 & 0.26 & 0.97 & 0.26 \\
    Trading profit                 & 0.33 & 0.29 & 0.95 & 0.33 \\
    \hline
  \end{tabular}
  \caption{Statistics for the equal-weighted portfolio on the \texttt{msci} universe.}
  \label{tab:sum-msci}
\end{table}

\newpage

\subsection{\texttt{msem} universe}

\begin{table}[!hbtp]
  \centering
  \begin{tabular}{|c|cc|cc|}
    \hline
    \textbf{Series} & \multicolumn{2}{|c|}{\texttt{lrg}} & \multicolumn{2}{|c|}{\texttt{sml}} \\
    & \textbf{Mean} & \textbf{St. dev.}  & \textbf{Mean} & \textbf{St. dev.} \\
    \hline
    Equal-weighted relative return &-0.67 & 4.31 & 1.00 & 4.94 \\
    Rebalancing premium            & 0.25 & 0.35 & 1.92 & 0.44 \\
    Trading profit                 & 0.38 & 0.53 & 1.84 & 0.74 \\
    \hline
  \end{tabular}
  \caption{Statistics for the equal-weighted portfolio on the \texttt{msem} universe.}
  \label{tab:sum-msem}
\end{table}

\section{Effect of transaction costs}

\begin{figure}[!htbp]
  \centering
  \includegraphics[width=0.49\textwidth]{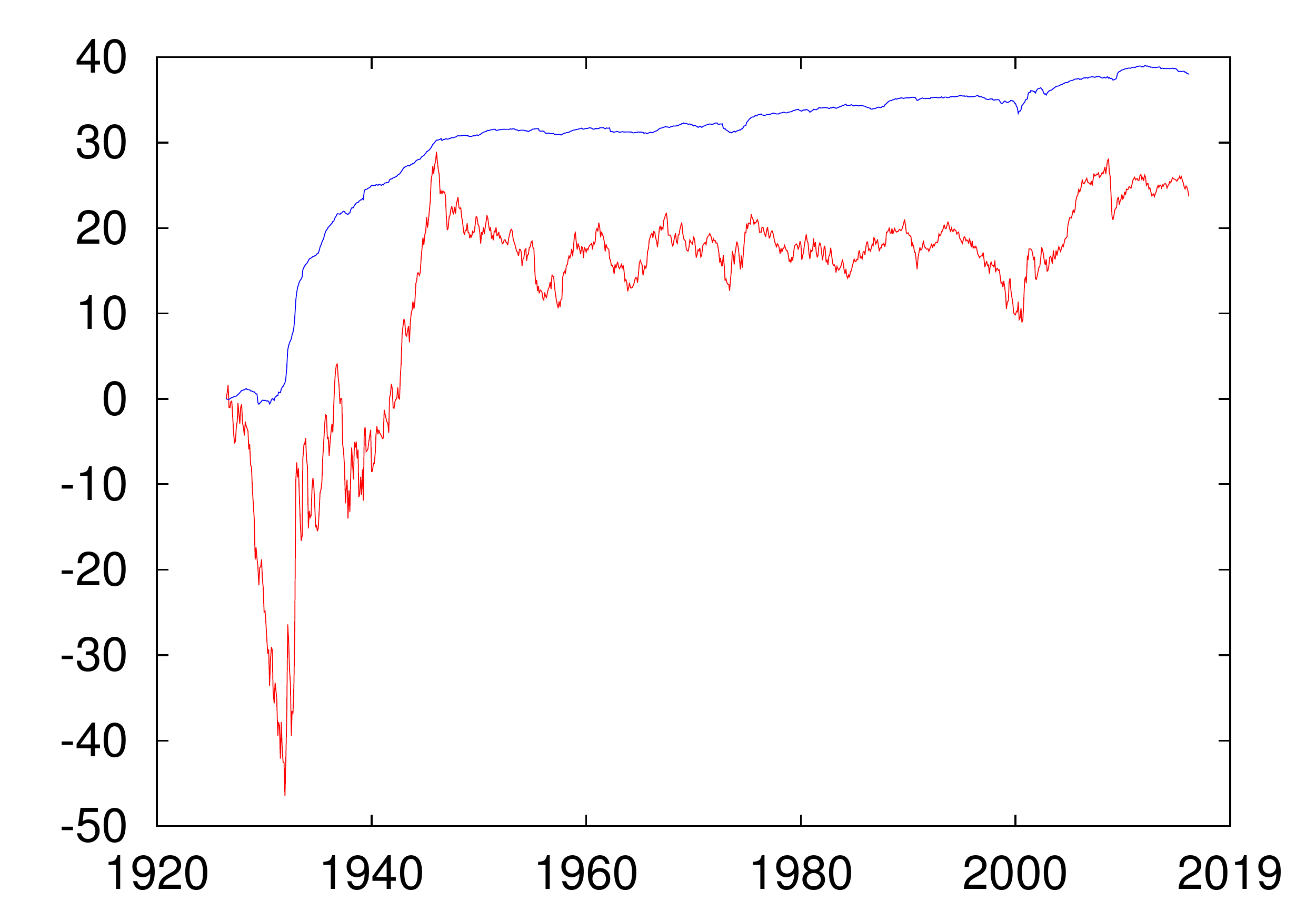}
  \includegraphics[width=0.49\textwidth]{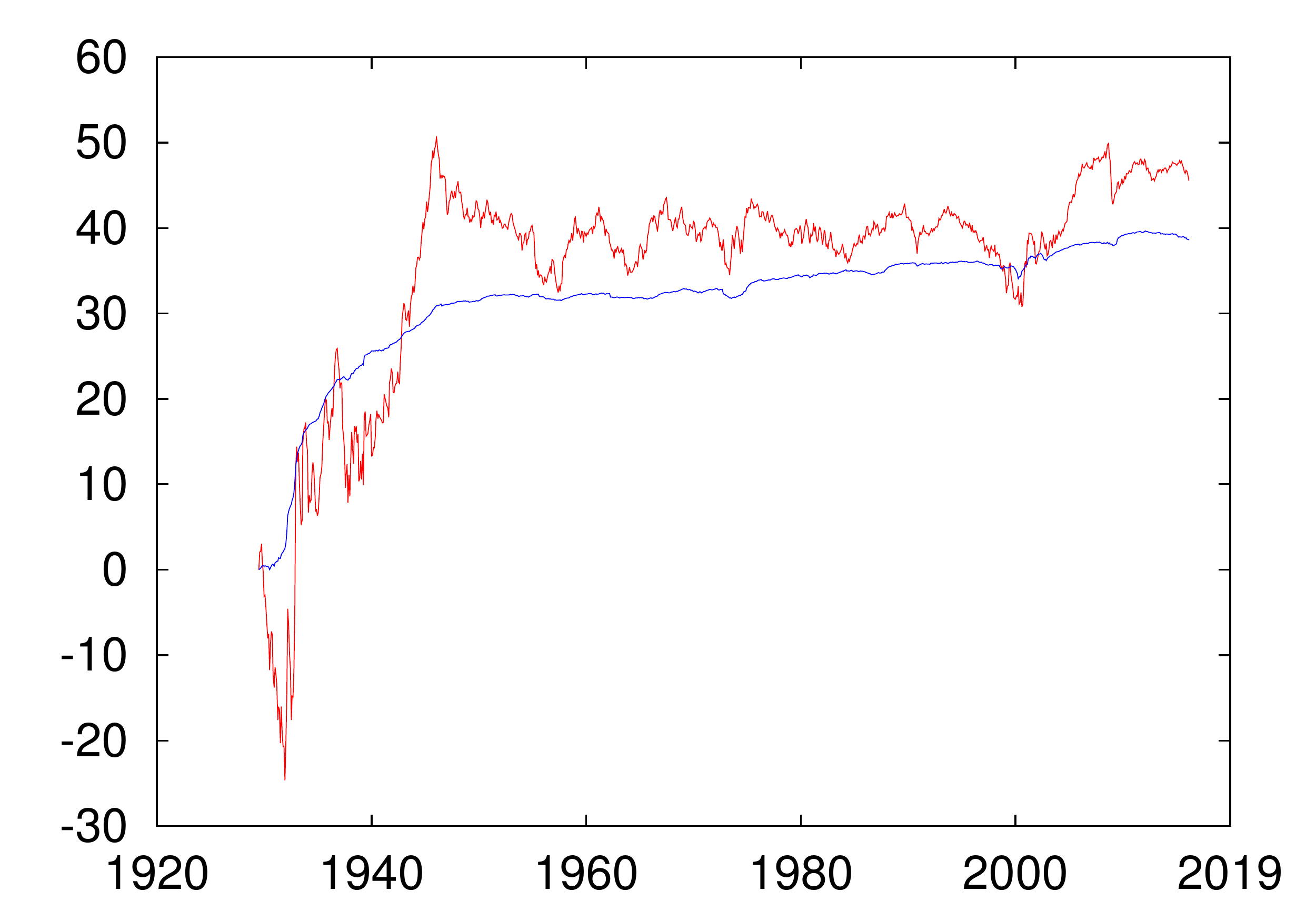}
  \caption{Cumulative performance of the equal-weighted portfolio for the top $n=\text{\texttt{sml}}$ stocks relative to the full market (red) and the trading-profit attribution (blue) for the \texttt{crsp} universe. The right-hand-side plots is redrawn starting in 1930.}
  \label{fig:pft-crsp-trd}
\end{figure}

\begin{figure}[!htbp]
  \centering
  \includegraphics[width=0.8\textwidth]{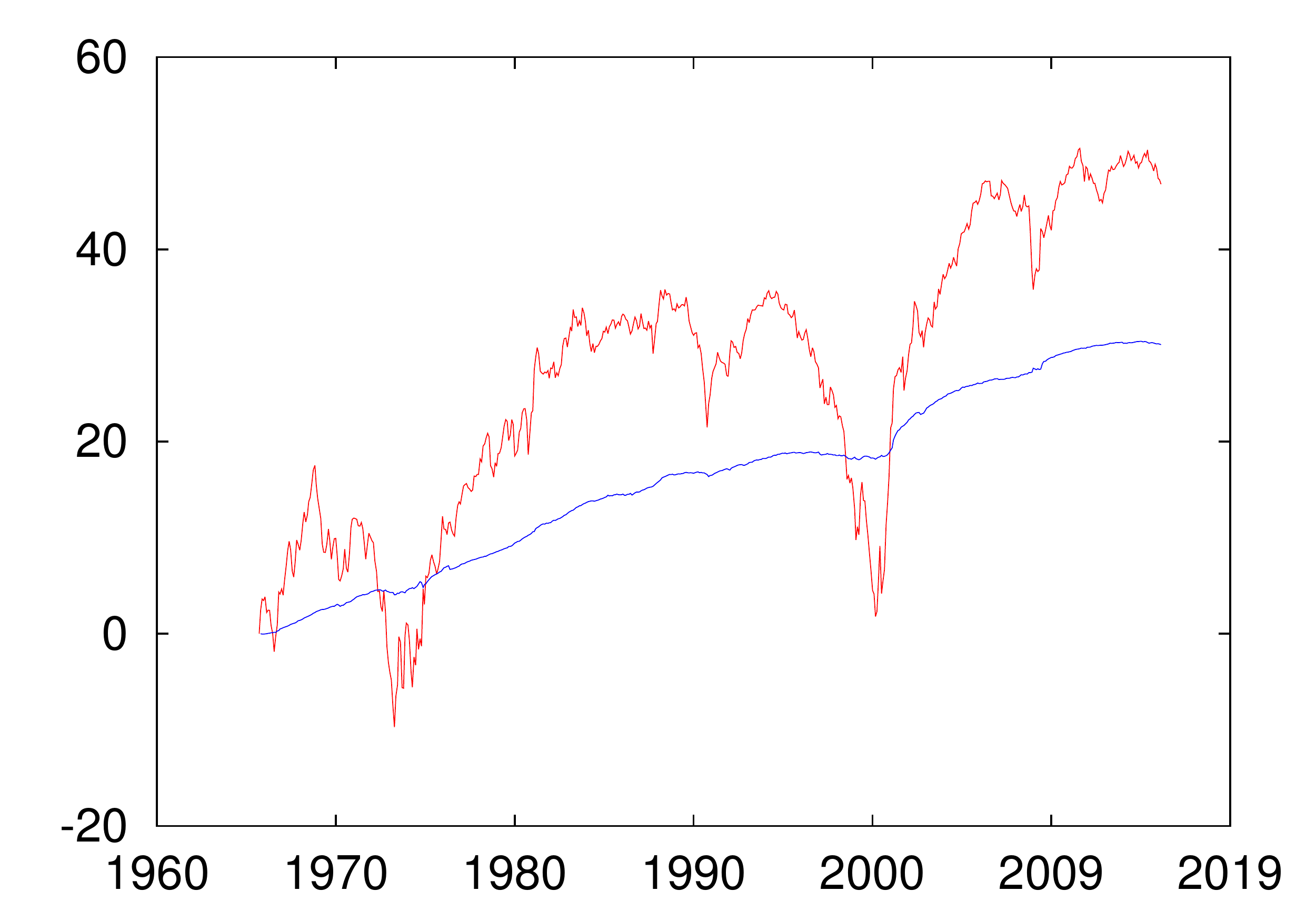}
  \caption{Same as \Rfig{pft-crsp-trd} for the \texttt{s500} universe.}
  \label{fig:pft-s500-trd}
\end{figure}

\begin{figure}[!htbp]
  \centering
  \includegraphics[width=0.45\textwidth]{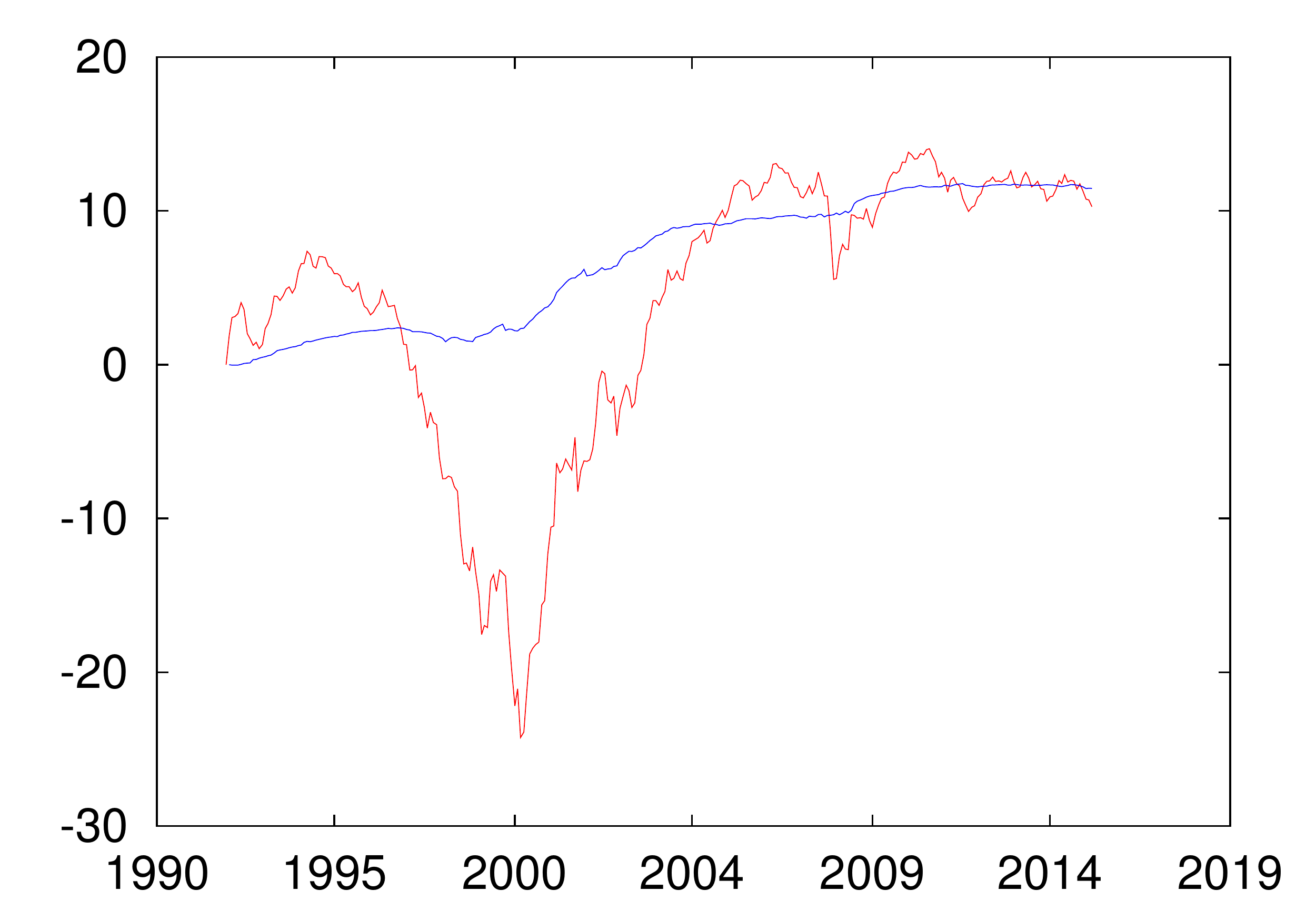}
  \caption{Same as \Rfig{pft-crsp-trd} for the \texttt{msci} universe.}
  \label{fig:pft-msci-trd}
\end{figure}

\begin{figure}[!htbp]
  \centering
  \includegraphics[width=0.45\textwidth]{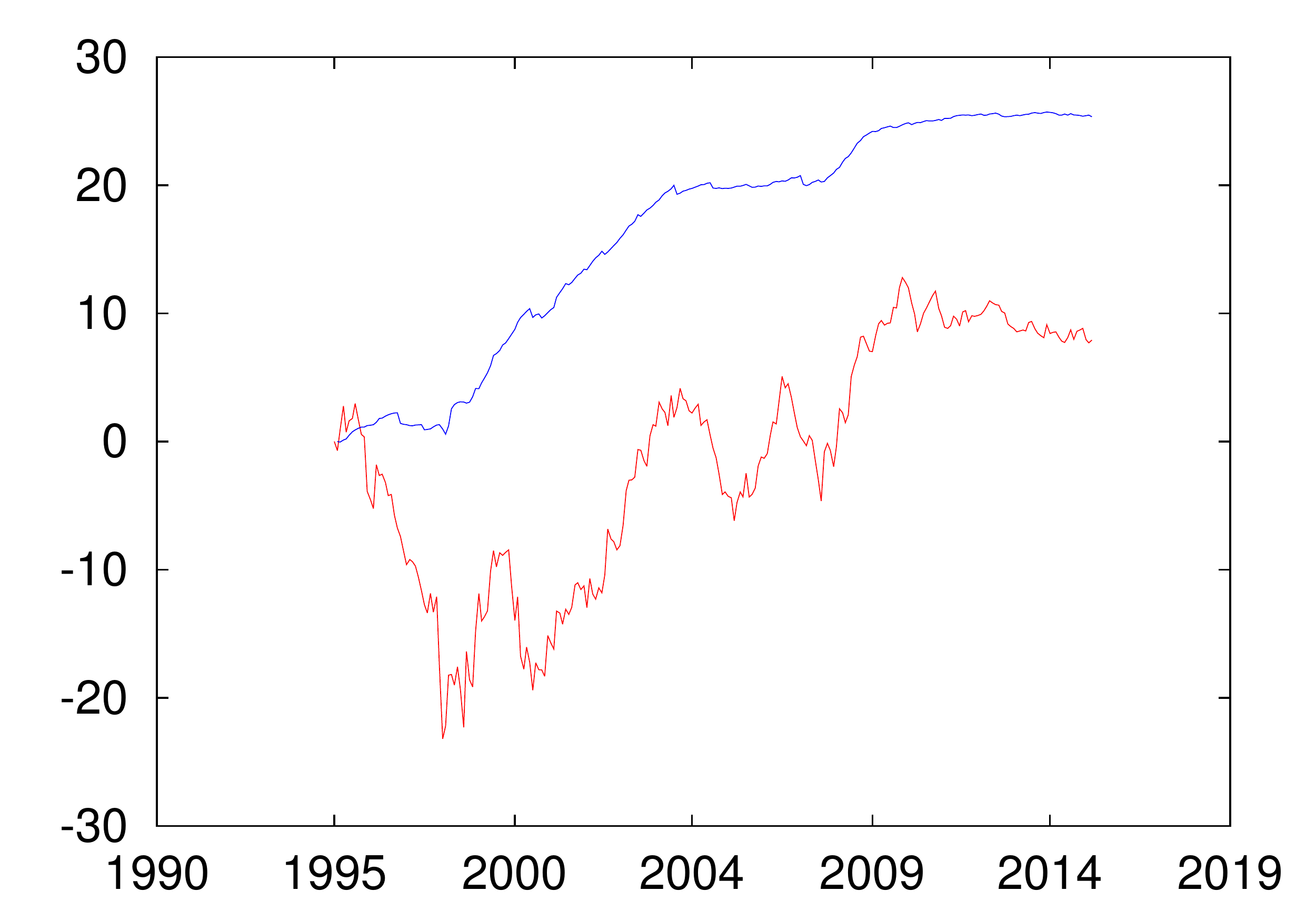}
  \caption{Same as \Rfig{pft-crsp-trd} for the \texttt{msem} universe.}
  \label{fig:pft-msem-trd}
\end{figure}

In this section, we consider the effect of accounting for transaction costs; as mentioned earlier, we assume transaction costs of 40~bps. We continue to assume that rebalancing occurs on a monthly frequency. The results of the simulations are shown in Tables~\ref{tab:sum-crsp-trd}--\ref{tab:sum-msem-trd}, as well as in Figures~\ref{fig:pft-crsp-trd}--\ref{fig:pft-msem-trd}. Since the effect of the universe magnitude was analyzed in the previous section, we fix the universe size. We also drop the SPT estimate of the rebalancing premium for simplicity.

\subsection{\texttt{crsp} universe}

\begin{table}[!hbtp]
  \centering
  \begin{tabular}{|c|cc|cc|}
    \hline
    \textbf{Series} & \multicolumn{2}{|c|}{\textbf{Full period}} & \multicolumn{2}{|c|}{\textbf{After 1930}} \\
    & \textbf{Mean}  & \textbf{Change}           & \textbf{Mean}  & \textbf{Change}          \\
    \hline
    Equal-weighted relative return & 0.27 & -0.47 & 0.53 & -0.47 \\
    Trading profit                 & 0.43 & -0.45 & 0.45 & -0.45 \\
    \hline
  \end{tabular}
  \caption{Statistics for the equal-weighted portfolio on the \texttt{crsp} universe. The columns labeled ``Change'' correspond to the difference with respect to the zero-transaction-costs case.}
  \label{tab:sum-crsp-trd}
\end{table}

\newpage

\subsection{\texttt{s500} universe}

\begin{table}[!hbtp]
  \centering
  \begin{tabular}{|c|cc|}
    \hline
    \textbf{Series} & \multicolumn{2}{|c|}{\texttt{sml}} \\
    & \textbf{Mean} & \textbf{Change} \\
    \hline
    Equal-weighted relative return  & 0.94 & -0.38 \\
    Trading profit                  & 0.60 & -0.39 \\
    \hline
  \end{tabular}
  \caption{Statistics for the equal-weighted portfolio on the \texttt{s500} universe.}
  \label{tab:sum-s500-trd}
\end{table}

\newpage

\subsection{\texttt{msci} universe}

\begin{table}[!hbtp]
  \centering
  \begin{tabular}{|c|cc|cc|}
    \hline
    \textbf{Series} & \multicolumn{2}{|c|}{\texttt{sml}} \\
    & \textbf{Mean} & \textbf{Change} \\
    \hline
    Equal-weighted relative return & 0.43 & -0.47 \\
    Trading profit                 & 0.48 & -0.47 \\
    \hline
  \end{tabular}
  \caption{Statistics for the equal-weighted portfolio on the \texttt{msci} universe.}
  \label{tab:sum-msci-trd}
\end{table}

\subsection{\texttt{msem} universe}

\begin{table}[!hbtp]
  \centering
  \begin{tabular}{|c|cc|cc|}
    \hline
    \textbf{Series} & \multicolumn{2}{|c|}{\texttt{sml}} \\
    & \textbf{Mean} & \textbf{Change} \\
    \hline
    Equal-weighted relative return & 0.38 & -0.62 \\
    Trading profit                 & 1.21 & -0.63 \\
    \hline
  \end{tabular}
  \caption{Statistics for the equal-weighted portfolio on the \texttt{msem} universe.}
  \label{tab:sum-msem-trd}
\end{table}

\newpage

\section{Effect of rebalancing frequency}

In this section, we consider the effect of varying the rebalancing frequency from monthly to quarterly and semiannual. To mitigate some of the complications due to calendar effects, we rebalance at an offset of two months, i.e.,~on the second month of each quarter (in the case of quarterly rebalancing), or on every February and August (in the case of semiannual rebalancing).

\begin{figure}[!htbp]
  \centering
  \includegraphics[width=0.44\textwidth]{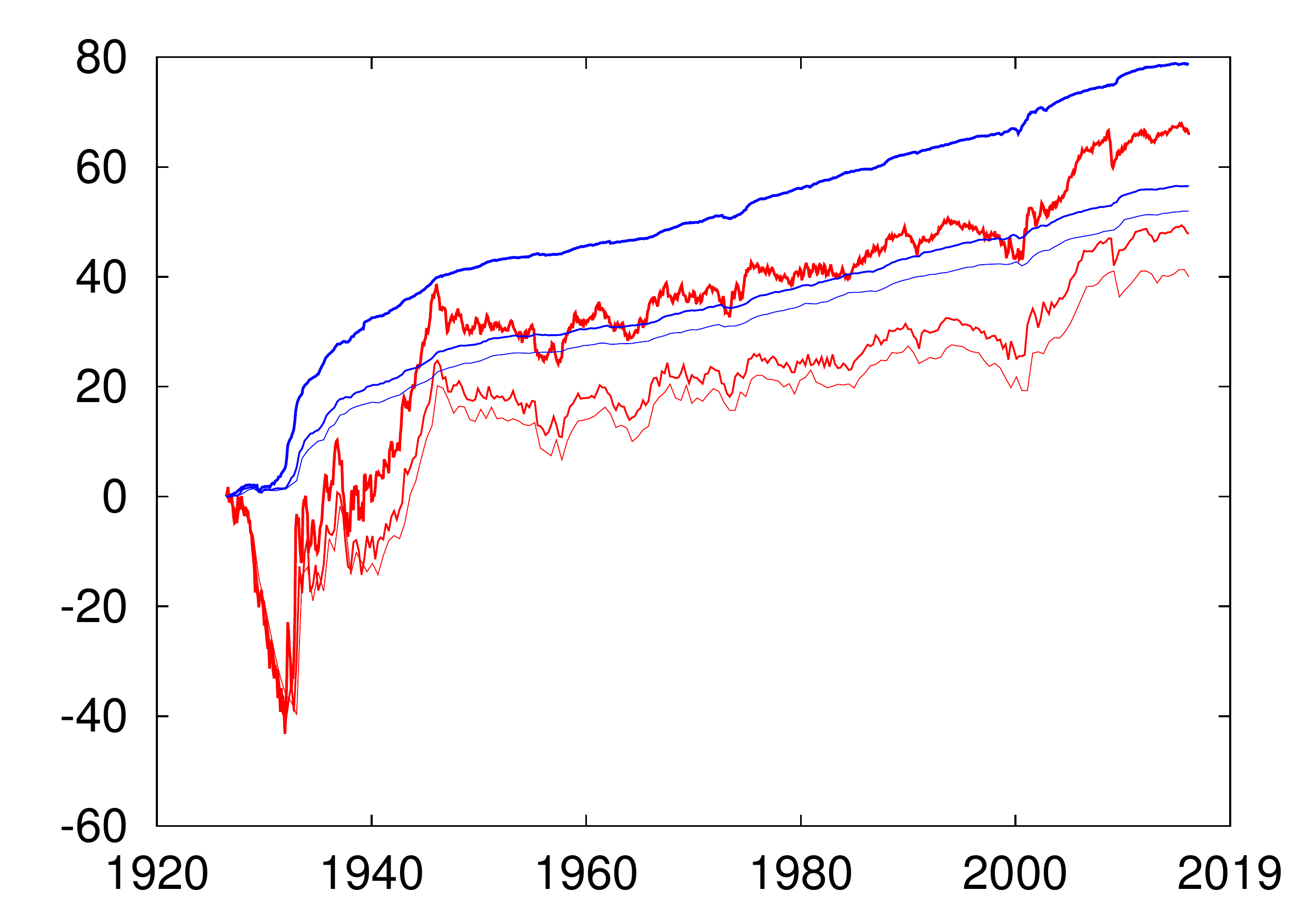}\qquad
  \includegraphics[width=0.44\textwidth]{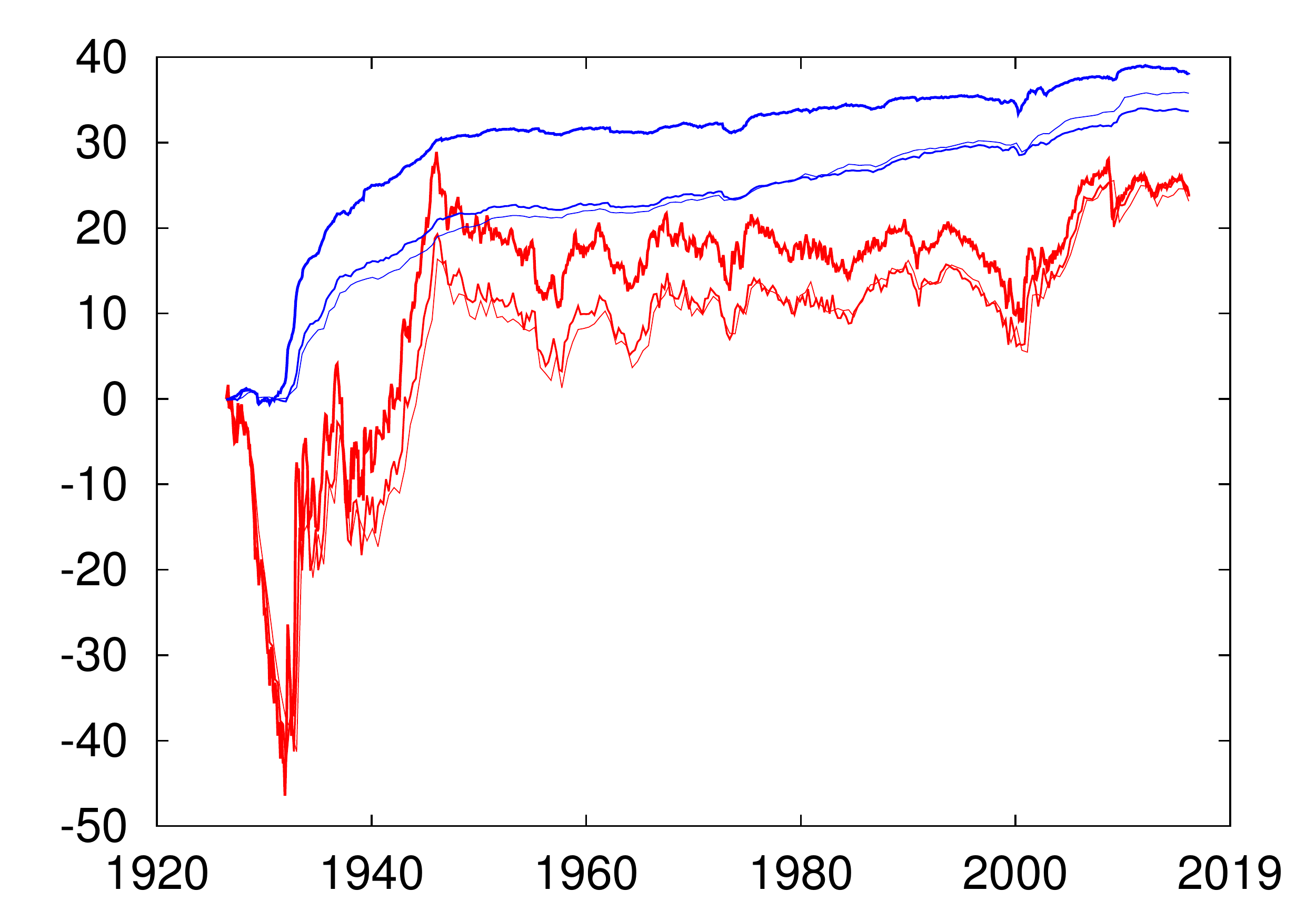}
  \caption{Cumulative performance of the equal-weighted portfolio for the top $n=\text{\texttt{sml}}$ stocks relative to the full market (red) and the trading-profit attribution (blue) for the \texttt{crsp} universe. The lines (ordered by decreasing thickness) correspond to monthly, quarterly and semiannual rebalancing. We assume that transaction costs are 0\,bps (left) and 40\,bps (right).}
  \label{fig:pft-crsp-reb}
\end{figure}

\begin{figure}[!htbp]
  \centering
  \includegraphics[width=0.31\textwidth]{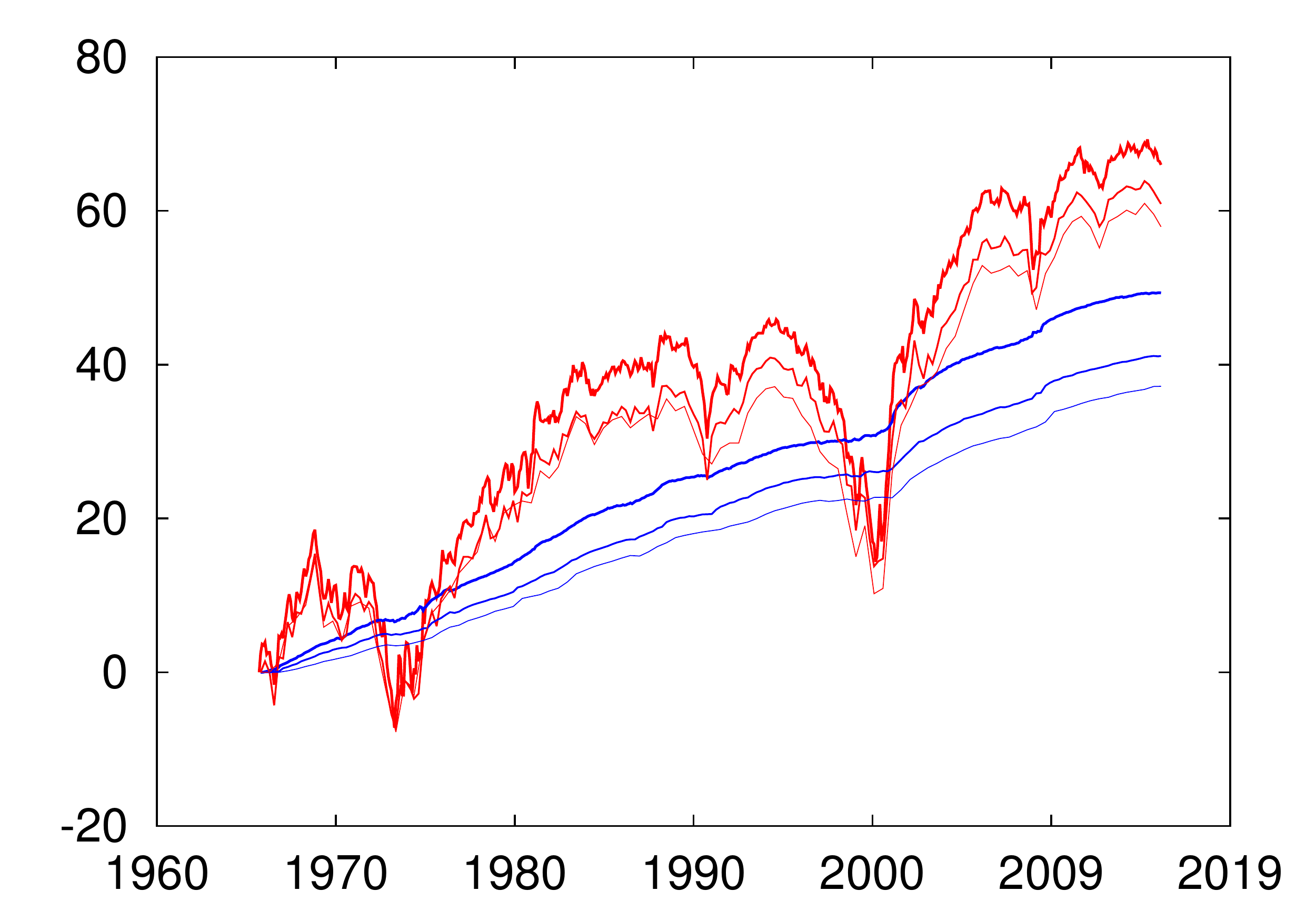}\qquad
  \includegraphics[width=0.31\textwidth]{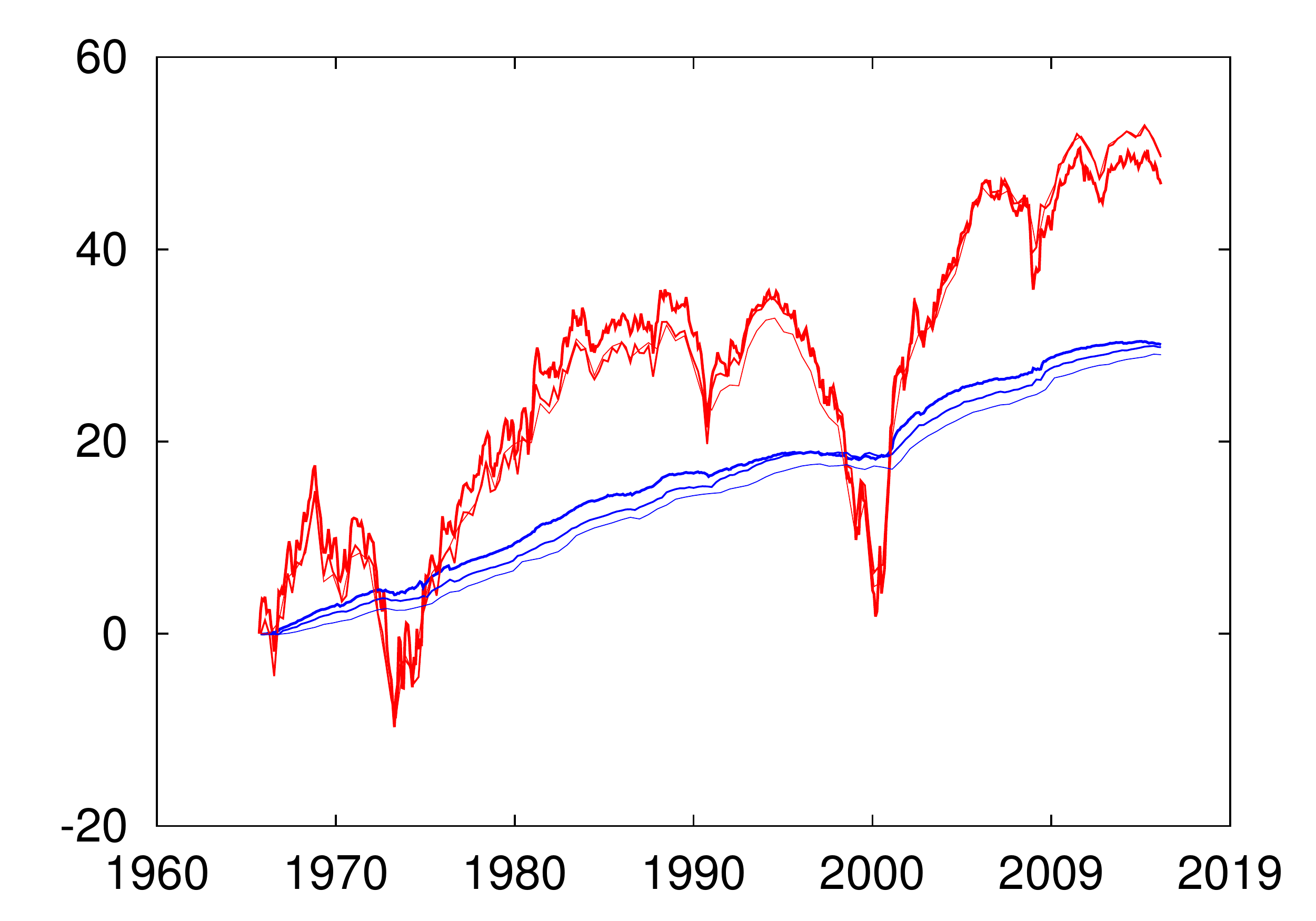}
  \caption{Same as \Rfig{pft-crsp-reb} for the \texttt{s500} universe.}
  \label{fig:pft-s500-reb}
\end{figure}

\begin{figure}[!htbp]
  \centering
  \includegraphics[width=0.31\textwidth]{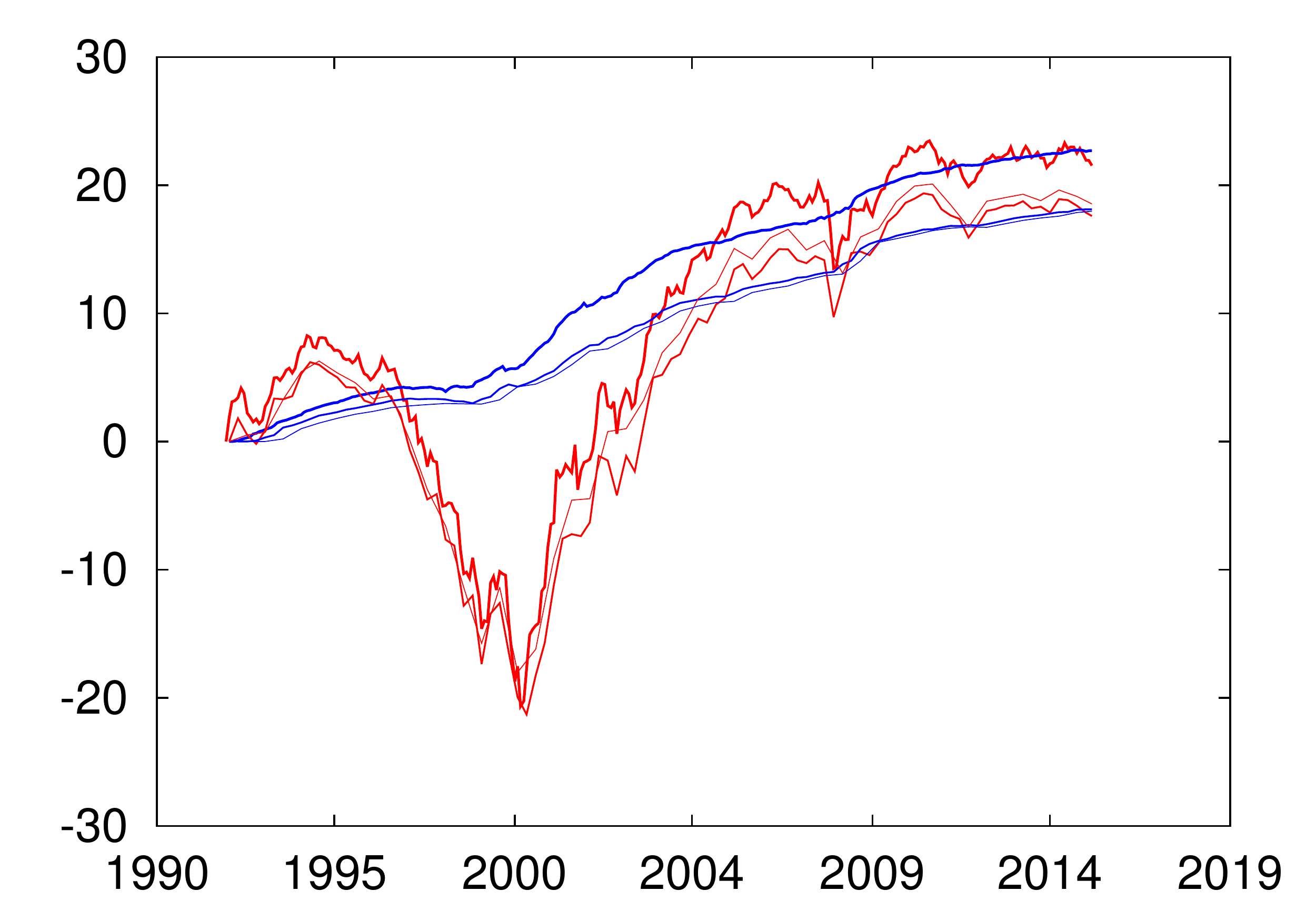}\qquad
  \includegraphics[width=0.31\textwidth]{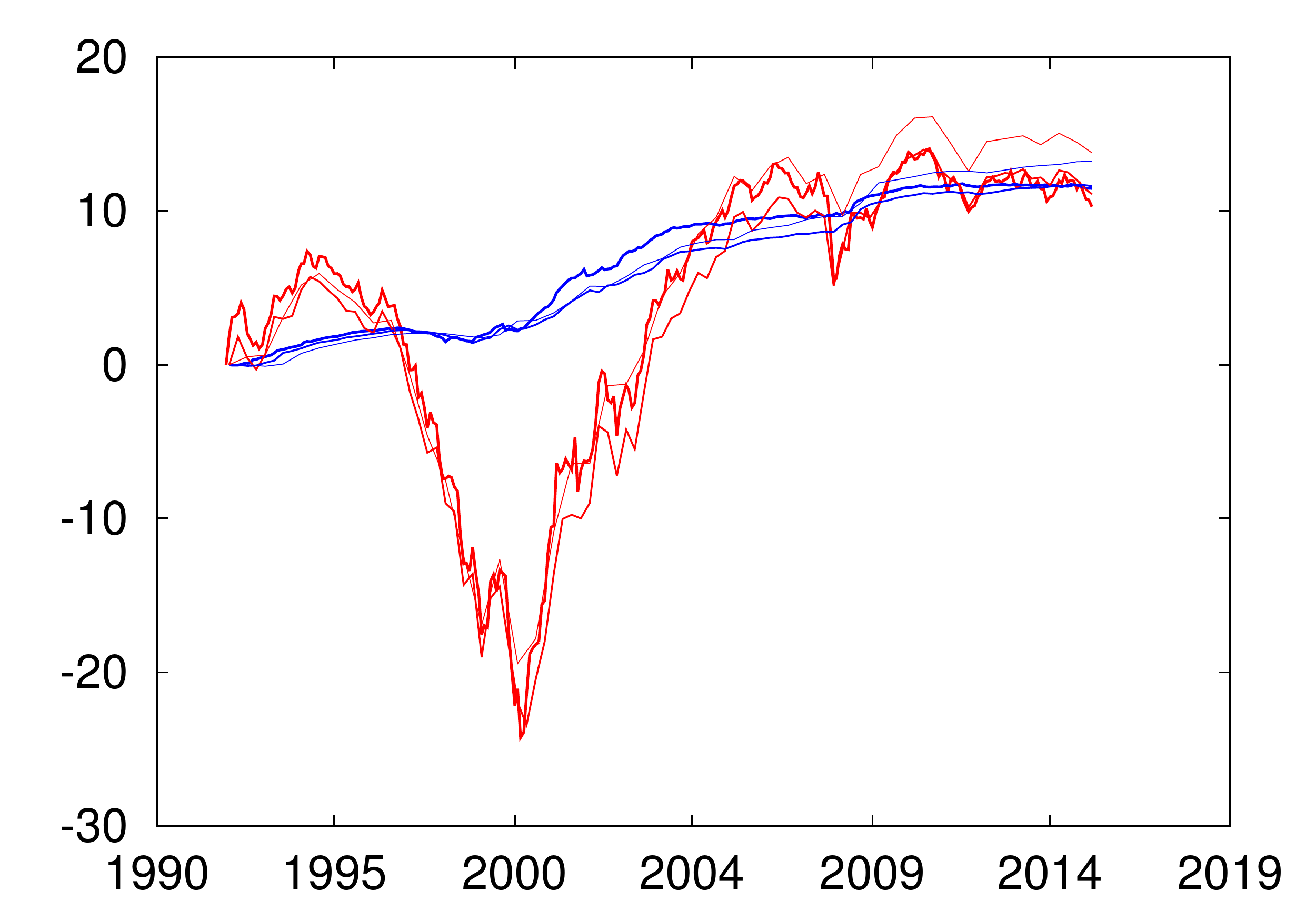}
  \caption{Same as \Rfig{pft-crsp-reb} for the \texttt{msci} universe.}
  \label{fig:pft-msci-reb}
\end{figure}

\begin{figure}[!htbp]
  \centering
  \includegraphics[width=0.47\textwidth]{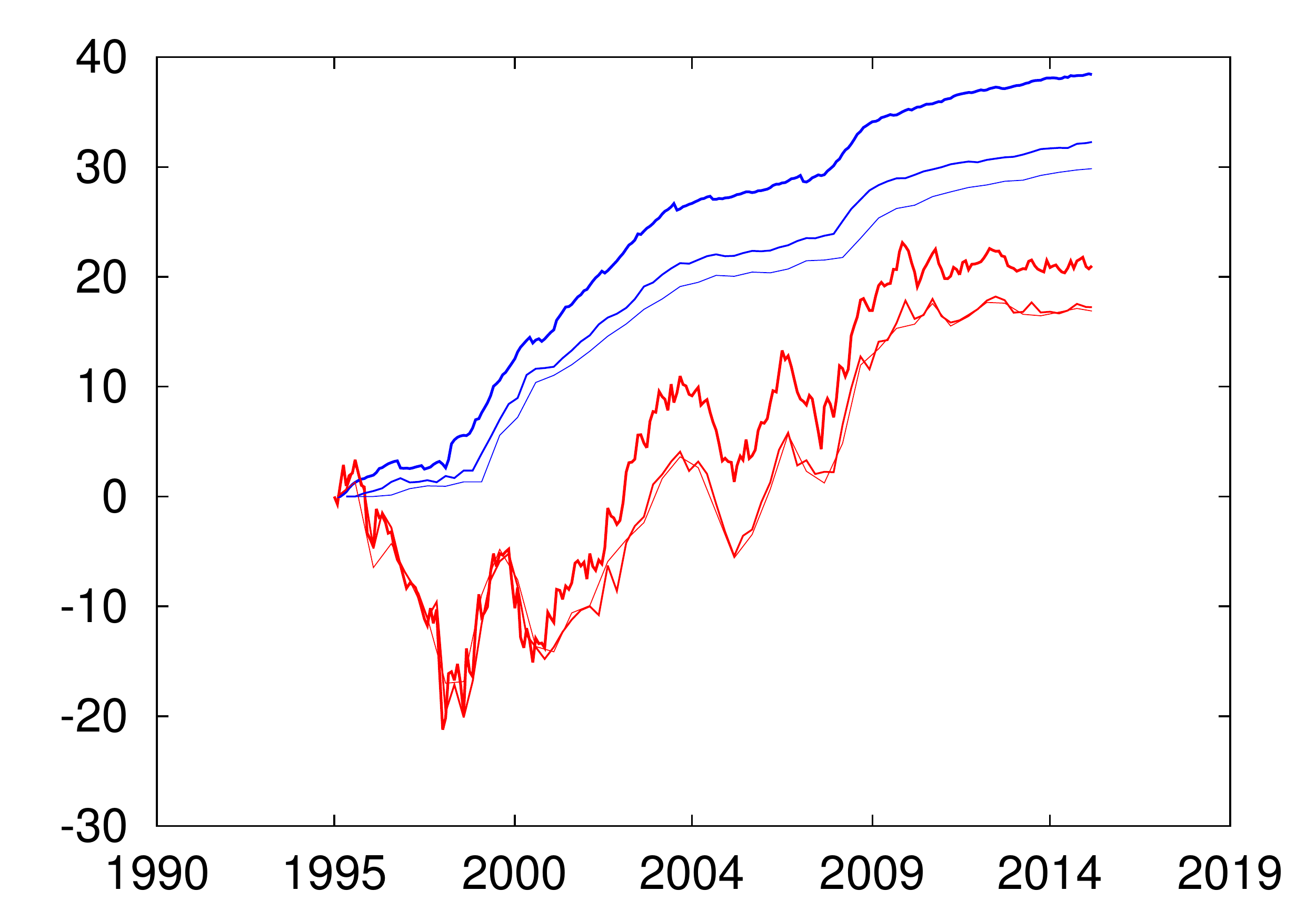}\qquad
  \includegraphics[width=0.47\textwidth]{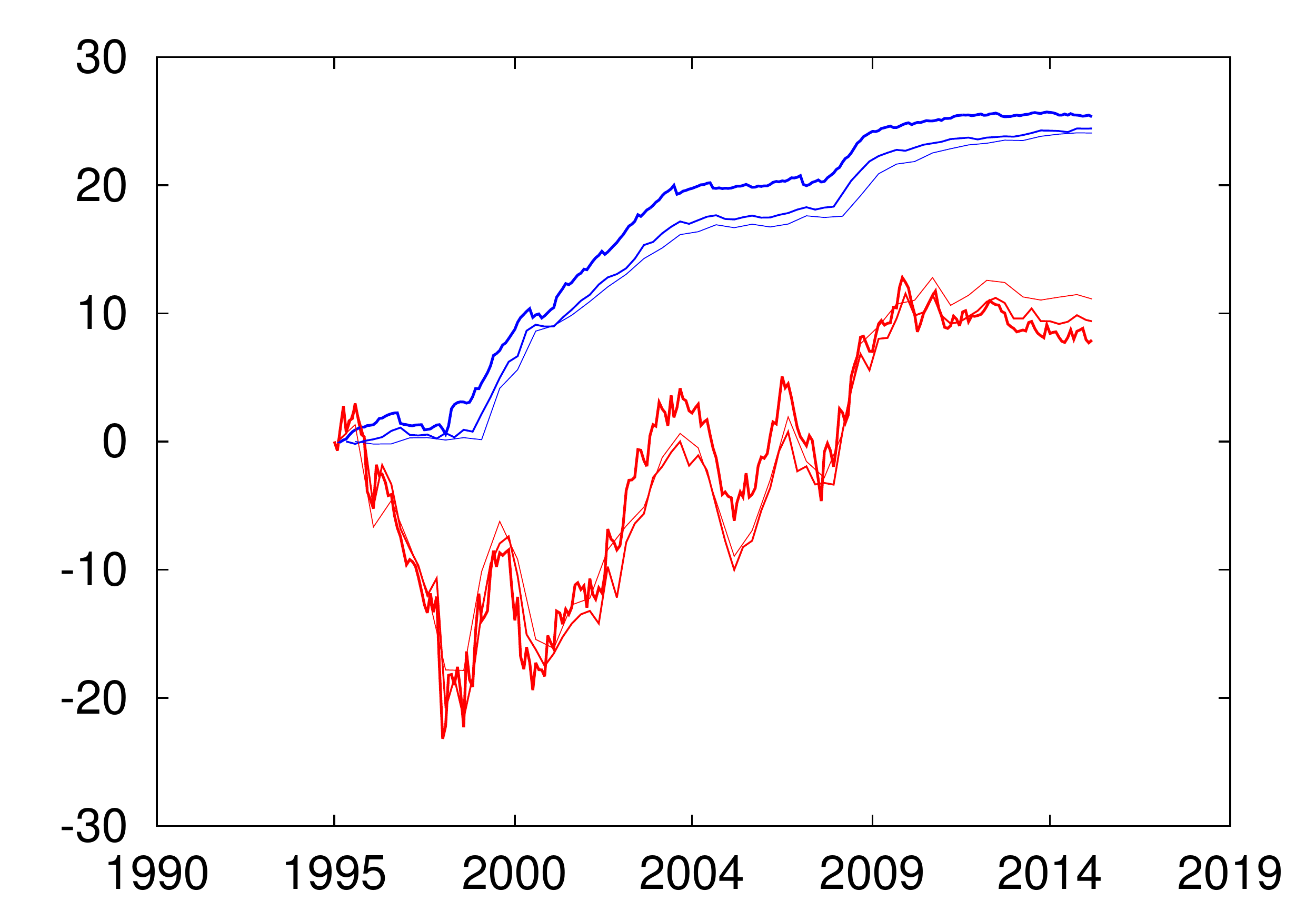}
  \caption{Same as \Rfig{pft-crsp-reb} for the \texttt{msem} universe.}
  \label{fig:pft-msem-reb}
\end{figure}

Rebalancing an equal-weighted portfolio less frequently misses opportunities to capture short-term volatility. On the other hand, it has the benefits of reducing both the turnover (and the associated transaction costs), as well as the performance drag due to reconstitution (as some of the securities revert to the top $n$ stocks over the medium term, which also affects the diversity exposure). In order to examine the two effects of transaction costs and rebalancing separately and together, we examine both cases of transaction costs of 0\,bps and 40\,bps. The results of the simulations are shown in Tables~\ref{tab:sum-crsp-reb}--\ref{tab:sum-msem-reb}, as well as in Figures~\ref{fig:pft-crsp-reb}--\ref{fig:pft-msem-reb}.
\subsection{\texttt{crsp} universe}

\begin{table}[!hbtp]
  \centering
      {\footnotesize
        \begin{tabular}{|c|c|cc|cc|}
          \hline
          \textbf{Series} & \multicolumn{1}{|c|}{\textbf{Monthly}} & \multicolumn{2}{|c|}{\textbf{Quarterly}} & \multicolumn{2}{|c|}{\textbf{Semiannual}} \\
          & \textbf{Mean} & \textbf{Mean} & \textbf{Change} & \textbf{Mean} & \textbf{Change} \\
          \hline
                          & \multicolumn{5}{|c|}{\textbf{Transaction costs of 0\,bps}} \\
          \hline
          Equal-weighted relative return & 0.74 & 0.54 & -0.20 & 0.45 & -0.29 \\
          Trading profit                 & 0.88 & 0.64 & -0.24 & 0.58 & -0.30 \\
          \hline
                          & \multicolumn{5}{|c|}{\textbf{Transaction costs of 40\,bps}} \\
          \hline
          Equal-weighted relative return & 0.27 & 0.27 &  0.00 & 0.26 & -0.01 \\
          Trading profit                 & 0.43 & 0.38 & -0.05 & 0.40 & -0.03 \\
          \hline
          Turnover                       & 59.2 & 33.4 & -25.9 & 23.7 & -35.6 \\
          \hline
        \end{tabular}
      }
      \caption{Statistics for the equal-weighted portfolio on the \texttt{crsp} universe.}
      \label{tab:sum-crsp-reb}
\end{table}

\newpage

\subsection{\texttt{s500} universe}

\begin{table}[!hbtp]
  \centering
      {\footnotesize
        \begin{tabular}{|c|c|cc|cc|}
          \hline
          \textbf{Series} & \multicolumn{1}{|c|}{\textbf{Monthly}} & \multicolumn{2}{|c|}{\textbf{Quarterly}} & \multicolumn{2}{|c|}{\textbf{Semiannual}} \\
          & \textbf{Mean} & \textbf{Mean} & \textbf{Change} & \textbf{Mean} & \textbf{Change} \\
          \hline
                          & \multicolumn{5}{|c|}{\textbf{Transaction costs of 0\,bps}} \\
          \hline
          Equal-weighted relative return & 1.32 & 1.22 & -0.10 & 1.16 & -0.16 \\
          Trading profit                 & 0.99 & 0.83 & -0.16 & 0.87 & -0.12 \\
          \hline
                          & \multicolumn{5}{|c|}{\textbf{Transaction costs of 40\,bps}} \\
          \hline
          Equal-weighted relative return & 0.94 & 0.99 & 0.05 & 1.00 &  0.06 \\
          Trading profit                 & 0.60 & 0.60 & 0.00 & 0.59 & -0.01 \\
          \hline
          Turnover                       & 48.1 & 28.3 & -19.8 & 20.4 & -27.7 \\
          \hline
        \end{tabular}
      }
      \caption{Statistics for the equal-weighted portfolio on the \texttt{s500} universe.}
      \label{tab:sum-s500-reb}
\end{table}

\subsection{\texttt{msci} universe}

\begin{table}[!hbtp]
  \centering
      {\footnotesize
        \begin{tabular}{|c|c|cc|cc|}
          \hline
          \textbf{Series} & \multicolumn{1}{|c|}{\textbf{Monthly}} & \multicolumn{2}{|c|}{\textbf{Quarterly}} & \multicolumn{2}{|c|}{\textbf{Semiannual}} \\
          & \textbf{Mean} & \textbf{Mean} & \textbf{Change} & \textbf{Mean} & \textbf{Change} \\
          \hline
                          & \multicolumn{5}{|c|}{\textbf{Transaction costs of 0\,bps}} \\
          \hline
          Equal-weighted relative return & 0.90 & 0.73 & -0.17 & 0.77 & -0.13 \\
          Trading profit                 & 0.95 & 0.76 & -0.19 & 0.77 & -0.18 \\
          \hline
                          & \multicolumn{5}{|c|}{\textbf{Transaction costs of 40\,bps}} \\
          \hline
          Equal-weighted relative return & 0.43 & 0.46 & 0.03 & 0.57 & 0.14 \\
          Trading profit                 & 0.48 & 0.49 & 0.01 & 0.56 & 0.08 \\
          \hline
          Turnover                       & 58.6 & 33.9 & -24.7 & 24.8 & -33.9 \\
          \hline
        \end{tabular}
      }
      \caption{Statistics for the equal-weighted portfolio on the \texttt{msci} universe.}
      \label{tab:sum-msci-reb}
\end{table}

\subsection{\texttt{msem} universe}

\begin{table}[!hbtp]
  \centering
      {\footnotesize
        \begin{tabular}{|c|c|cc|cc|}
          \hline
          \textbf{Series} & \multicolumn{1}{|c|}{\textbf{Monthly}} & \multicolumn{2}{|c|}{\textbf{Quarterly}} & \multicolumn{2}{|c|}{\textbf{Semiannual}} \\
          & \textbf{Mean} & \textbf{Mean} & \textbf{Change} & \textbf{Mean} & \textbf{Change} \\
          \hline
                          & \multicolumn{5}{|c|}{\textbf{Transaction costs of 0\,bps}} \\
          \hline
          Equal-weighted relative return & 1.00 & 0.82 & -0.18 & 0.80 & -0.20 \\
          Trading profit                 & 1.84 & 1.56 & -0.28 & 1.46 & -0.38 \\
          \hline
                          & \multicolumn{5}{|c|}{\textbf{Transaction costs of 40\,bps}} \\
          \hline
          Equal-weighted relative return & 0.38 & 0.45 &  0.07 & 0.53 &  0.15 \\
          Trading profit                 & 1.21 & 1.18 & -0.03 & 1.17 & -0.04 \\
          \hline
          Turnover                       & 77.8 & 46.7 & -31.2 & 34.3 & -43.5 \\
          \hline
        \end{tabular}
      }
      \caption{Statistics for the equal-weighted portfolio on the \texttt{msem} universe.}
      \label{tab:sum-msem-reb}
\end{table}

\section{Conclusion}

The trading-profit attribution methodology leads to results that are consistent with the long-term relative outperformance of the equal-weighted portfolio, and provides an alternate method for estimating the rebalancing premium without the requirement of waiting for the market cycle regarding the size exposure to complete. Furthermore, the behavior of the trading-profit contribution to the equal-weighted portfolio performance reflects closely the impact of various effects, such as changing the magnitude of the universe, the transaction costs, or the rebalancing frequency.

All these experiments furnish further strong support for the claim that the underlying cause for the outperformance of the equal-weighted portfolio relative to the cap-weighted market is the systematic capture of trading profit through the process of rebalancing to the slowly-varying\footnote{Even though the equal-weighted portfolio has constant target weights between reconstitutions, each reconstitution event results in a change of the target weights for those securities that drop out of, or are newly included into, the investable universe.} target weights.


\begin{thebibliography}{9}
\bibitem{RP} R.~Yasenchak, V.~Papathanakos, \emph{Measuring the Rebalancing Premium: A New Portfolio Attribution Framework}, white paper, \texttt{www.intechjanus.com}, 2015.
\bibitem{BPW} A.~Banner, V.~Papathanakos, P.~Whitman, \emph{Rebalancing Act: Defying the Laws of Traditional Finance Theory, Low-Volatility Portfolios Outperform More Volatile Peers.}, Institutional Investor, October 2012.
\bibitem{F} E.~R.~Fernholz, \emph{Stochastic Portfolio Theory}, Springer, 2002.
\bibitem{FK} E.~R.~Fernholz, I.~Karatzas, \emph{Stochastic Portfolio Theory:  An Overview}, Handbook of Numerical Analysis, volume ``Mathematical Modeling and Numerical Methods in Finance'' (A. Bensoussan, ed.), pp. 89--168, \url{http://www.math.columbia.edu/~ik/FernKarSPT.pdf}, 2009.
\bibitem{C} Database d6z201512 of the Center for Research in Security Prices (CRSP), The University of Chicago Booth School of Business, 2016.
\bibitem{S} S\&P Dow Jones Indices LLC, \texttt{us.spindices.com}, 2016.
\bibitem{M} MSCI Inc, \texttt{www.msci.com}, 2016.
\end{thebibliography}
\end{document}